\documentclass[aps,prl,floatfix,twocolumn,showpacs]{revtex4}%
\usepackage{amssymb}
\usepackage{amsmath}
\usepackage{graphicx}
\usepackage{amsfonts}

\begin{document}
\title{Positive cross-correlations due to Dynamical Channel-Blockade in a
three-terminal quantum dot}
\author{A. Cottet, W. Belzig, and C. Bruder}
\affiliation{Department of Physics and Astronomy, University of Basel, Klingelbergstrasse
82, 4056 Basel, Switzerland}

\begin{abstract}
We investigate current fluctuations in a three-terminal quantum dot in the
sequential tunneling regime. In the voltage-bias configuration chosen here,
the circuit is operated like a beam splitter, i.e. one lead is used as an
input and the other two as outputs. In the limit where a double occupancy of
the dot is not possible, a super-Poissonian Fano factor of the current in the
input lead and positive cross-correlations between the current fluctuations in
the two output leads can be obtained, due to dynamical channel-blockade. When
a single orbital of the dot transports current, this effect can be obtained by
lifting the spin-degeneracy of the circuit with ferromagnetic leads or with a
magnetic field. When several orbitals participate in the electronic
conduction, lifting spin-degeneracy is not necessary. In all cases, we show
that a super-Poissonian Fano factor for the input current is not equivalent to
positive cross-correlations between the outputs. We identify the conditions
for obtaining these two effects and discuss possible experimental realizations.

\end{abstract}
\date{\today}
\maketitle

%73.23.-b Electronic transport in mesoscopic systems
%72.70.+m Noise processes and phenomena
%72.25.Mk Spin transport through interfaces
%72.25.Rb Spin relaxation and scattering

\section{I. INTRODUCTION}

The study of current noise in mesoscopic circuits has become a central
subfield of mesoscopic physics because it allows to access informations not
available through measurements of the average currents (for reviews, see
Refs.~\cite{blanter:00,nazarov:03}). Current fluctuations can first be probed
through the auto-correlations of the current fluctuations in one branch of the
circuit. For conductors with open channels, the fermionic statistics of
electrons result in a suppression of these auto-correlations below the Poisson
limit \cite{khlus:87, lesovik:89, buettiker:90}. In a multi-terminal circuit,
current fluctuations can also be probed through the cross-correlations between
two different branches. B\"{u}ttiker has shown that \textit{in a
non-interacting electronic circuit, the zero-frequency current
cross-correlations are always negative provided the leads of the circuit are
thermal reservoirs maintained at constant\ voltage-potentials }
\cite{buettiker:92}. On the experimental side, negative cross-correlations
have been measured very recently by Henny \textit{et al.} \cite{henny:99} and
Oliver \textit{et al.} \cite{oliver:99} in mesoscopic beam splitters.
Oberholzer \textit{et al.} have shown how the cross-correlations vanish in the
classical limit \cite{oberholzer:00}.

Up to now, positive cross-correlations have never been measured in electronic
circuits. However, nothing forbids to reverse the sign of cross-correlations
if a hypothesis of B\"{u}ttiker's proof is not fulfilled (see Ref.
\cite{buettiker:03-book} for a recent review). First, it has been shown
theoretically that positive cross-correlations can be obtained in an
electronic circuit by relaxing the hypotheses of B\"{u}ttiker regarding the
leads, for instance by taking one of the leads superconducting
\cite{p1,p2,p3,p4,p5,p5b,p6,p7,p8,p9,p10,p11}, or by using leads with an
imperfect \cite{texier:00} or time-dependent \cite{Gattobigio} voltage bias.
Positive cross-correlations are also expected at finite frequencies, due to
the plasmonic screening currents existing in capacitive circuits
\cite{martin:00,buettiker:03-book}. It follows from B\"{u}ttiker's work that
obtaining positive cross-correlations at zero frequency without modifying the
assumptions on the leads requires to have interactions \textit{inside} the
device. Safi \textit{et al.} have considered a two dimensional electron gas in
the fractional quantum Hall regime, described by a chiral Luttinger liquid
theory \cite{safi:01}. Zero-frequency positive cross-correlations can be
obtained in this system in the limit of small filling factors, where the
excitations of the chiral Luttinger liquid take a bosonic character. This
leaves open the question whether interactions localized inside the beam
splitter can lead to zero-frequency positive cross-correlations even for a
\textit{normal fermionic circuit.}

Current correlations in a single quantum dot have been studied in the
sequential tunneling limit \cite{setnoise,bagrets:02,birk:95,bulka:99}, in the
cotunneling regime \cite{Sukhorukov,Averin} and in the Kondo regime
\cite{Kondo}. In the (spin-degenerate) sequential tunneling limit, a
sub-Poissonian Fano factor has been found for some two-terminal cases
\cite{setnoise,bagrets:02,birk:95}, and, for the three-terminal case,
cross-correlations are expected to be always negative when the intrinsic level
spacing $\Delta E$ of the dot is much smaller than temperature
\cite{bagrets:02}. However, a super-Poissonian Fano factor has been predicted
for a two-terminal quantum dot with $\Delta E\gg k_{B}T$ connected to
ferromagnetic leads \cite{bulka:99}. In the cotunneling regime, a
super-Poissonian Fano factor can be obtained in the two-terminal case
\cite{Sukhorukov}. The extent to which this would lead to positive
cross-correlations for a three-terminal quantum dot was not clear.

In this article, we consider a three-terminal quantum dot with $\Delta E\gg
k_{B}T$, operated as a beam splitter: one contact acts as source and the other
two as drains. In earlier papers, we have proposed two different ways to
obtain zero-frequency positive cross-correlations in this circuit, in the
sequential tunneling limit. We have assumed that only one orbital of the dot,
i.e. one single-particle level, transports current. Both methods rely on
lifting spin-degeneracy, either by using ferromagnetic leads \cite{audrey:03},
or by using paramagnetic leads but placing the dot in a magnetic field
\cite{audrey:04}. Note that in these works, the leads are biased with constant
voltages and modeled as non-interacting Fermi gases. Then, with respect to
B\"{u}ttiker's proof, only the hypothesis of the absence of interactions
inside the device itself is relaxed. Moreover, in contrast to the system
studied in \cite{safi:01}, excitations inside the device remain purely fermionic.

We provide here a detailed analysis of the physical origin of the positive
cross-correlations found in a three-terminal interacting quantum dot. The
essential ingredient is the existence of Coulomb interactions on the dot.
(Note that in a spin valve connected to ferromagnetic leads, in which there
are no charging effects, the cross-correlations where found to be negative
\cite{belzig:03}). Here, we assume that Coulomb interactions prevent a double
occupancy of the dot. In the limit were only one orbital level of the dot
transports current, the mechanism responsible for positive cross-correlations
is \textit{dynamical spin-blockade}. Simply speaking, up- and down-spins
tunnel through the dot with different rates. The spins which tunnel with a
lower rate modulate the transport through the opposite spin-channel, which
leads to a bunching of tunneling events. We consider both the Fano factor in
the input lead, called input Fano factor, and the cross-correlations between
the two output leads, called output cross-correlations. We show that a
super-Poissonian input Fano factor is not equivalent to positive output
cross-correlations and identify the conditions to obtain these effects. We
furthermore show that there is a direct mapping between the above case of a
non spin-degenerate quantum dot with a single orbital-level transporting
current and the case of a spin-degenerate quantum dot with two orbital levels
transporting current. This mapping implies that the result of
\cite{bagrets:02} cannot be generalized to $\Delta E\gg k_{B}T$:
cross-correlations are not always negative for a spin-degenerate
three-terminal quantum dot. More generally, this result provides the evidence
that\textit{\ lifting spin-degeneracy is not necessary }for obtaining
zero-frequency positive cross-correlations due to interactions inside a beam
splitter device, even for a normal fermionic circuit with a perfect voltage
bias\textit{. }In this spin-degenerate case, positive cross-correlations stem
from\ the partial blockade of an electronic channel by another one, thus we
propose to call this effect: \textit{dynamical channel-blockade.}

The present article is organized as follows. Section II develops the
mathematical description valid for the one-orbital problem. This one-orbital
problem is analyzed for two different configurations. First, the case of
ferromagnetic leads and zero magnetic field is treated in Section III.
Secondly, the case of a Zeeman splitting created by a magnetic field is
treated in Section IV. In Section V, we show how to map the two-orbital
spin-degenerate problem onto the one-orbital problem.

\section{II. MODEL\ AND\ GENERAL\ DESCRIPTION FOR\ THE\ ONE-ORBITAL\ CASE}

\subsection{A. Model}

\begin{figure}[h]
\includegraphics[width=0.5\linewidth,angle=-90,clip]{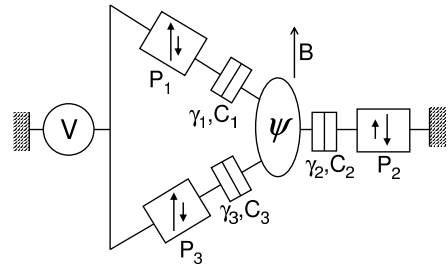}\caption{Electrical
diagram of a quantum dot connected to three leads $i\in\{1,2,3\}$ with
collinear magnetic polarizations $P_{i}$, through tunnel junctions with net
tunneling rates $\gamma_{i}$ and capacitances $C_{i}$. A bias voltage $V$ is
applied to leads $1$ and $3$; lead $2$ is connected to ground. A magnetic
field $B$ collinear to the lead polarizations is applied to the dot.}%
\label{Pict1}%
\end{figure}

We consider a quantum dot connected to three leads $i\in\{1,2,3\}$, through
tunnel junctions with capacitances $C_{i}$ and net spin-independent tunneling
rates $\gamma_{i}$ (Fig. \ref{Pict1}). The leads are magnetically polarized in
collinear directions. We also assume that the dot is subject to a magnetic
field $B$ collinear to the lead polarizations. A voltage bias $V$ is applied
to leads $1$ and $3$ whereas lead $2$ is connected to ground. The voltage $V$
is considered as positive, such that it is energetically more favorable for
electrons to go from the input electrode 2 to the output electrodes 1 or 3
than in the opposite direction. In this section, we also assume that
\begin{equation}
k_{B}T,\mu_{B}B,eV\ll E_{C},\Delta E\,\text{,} \label{eq:parameters}%
\end{equation}
where the charging energy $E_{C}=e^{2}/2C$ of the dot depends on $C=\sum
_{i}C_{i}$. According to (\ref{eq:parameters}), only one orbital level of the
dot, with energy $E_{0}$, needs to be taken into account to describe the
current transport, and this level cannot be doubly occupied. In this
situation, there are three possible states $\psi$ for the dot: either empty
i.e. $\psi=0$, or occupied with one electron with spin $\sigma\in
\{\uparrow,\downarrow\}$ i.e. $\psi=\sigma$. The magnetic field $B$ induces a
Zeeman splitting of the level according to $E_{\downarrow(\uparrow)}%
=E_{0}+(-)g\mu_{B}B/2$, where $\mu_{B}=e\hbar/2m$ is the Bohr magneton. In
this article, we will assume $B\geq0$, i.e. the up-spin level is energetically
lower than the down-spin level in the presence of a magnetic field. The
collinear magnetic polarizations $P_{j}$ of the leads are taken into account
by using \textit{spin-dependent} tunneling rates $\gamma_{j\uparrow}%
=\gamma_{j}(1+P_{j})$ and $\gamma_{j\downarrow}=\gamma_{j}(1-P_{j})$. In a
simple model, the spin-dependence is a consequence of the different densities
of states for electrons with up and down spins in the leads \cite{julliere:75}%
. The rate for an electron to tunnel on/off the dot ($\epsilon=+/-1$) through
junction $j$ is then given by
\begin{equation}
\Gamma_{j\sigma}^{\epsilon}=\gamma_{j\sigma}/(1+\exp[\epsilon(E_{\sigma
}-eV_{j})/k_{B}T])\text{ ,} \label{rates}%
\end{equation}
where $V_{1}=V_{3}=-C_{2}V/C$ and $V_{2}=(C_{1}+C_{3})V/C$. Here, we took the
Fermi energy $E_{F}=0$ for lead 2 as a reference. On the dot, there can be
spin-flip scattering, due for instance to spin-orbit coupling or to magnetic
impurities. According to the detailed balance rule, we write the spin-flip
rates as
\[
\Gamma_{\uparrow\downarrow}=\gamma_{sf}\exp(+\frac{g\mu_{B}B}{2k_{B}T})
\]
for the $\downarrow\longrightarrow\uparrow$ transition and
\[
\Gamma_{\downarrow\uparrow}=\gamma_{sf}\exp(-\frac{g\mu_{B}B}{2k_{B}T})
\]
for the $\uparrow\longrightarrow\downarrow$ transition.

\subsection{B. Master equation treatment}

In the sequential-tunneling limit $\hbar\gamma_{j\sigma}\ll k_{B}T$,
electronic transport through the dot can be described by the master equation
\cite{setnoise}
\begin{equation}
\frac{d}{dt}\left[
\begin{array}
[c]{c}%
p_{\uparrow}(t)\\
p_{\downarrow}(t)\\
p_{0}(t)
\end{array}
\right]  =\widehat{M}\left[
\begin{array}
[c]{c}%
p_{\uparrow}(t)\\
p_{\downarrow}(t)\\
p_{0}(t)
\end{array}
\right]  \;, \label{MasterEquation}%
\end{equation}
where $p_{\psi}(t)$, $\psi\in\{\uparrow,\downarrow,0\}$, is the instantaneous
occupation probability of state $\psi$ at time $t$, and where
\begin{equation}
\widehat{M}=\left[
\begin{array}
[c]{ccc}%
-\Gamma_{\uparrow}^{-}-\Gamma_{\downarrow\uparrow} & \Gamma_{\uparrow
\downarrow} & \Gamma_{\uparrow}^{+}\\
\Gamma_{\downarrow\uparrow} & -\Gamma_{\downarrow}^{-}-\Gamma_{\uparrow
\downarrow} & \Gamma_{\downarrow}^{+}\\
\Gamma_{\uparrow}^{-} & \Gamma_{\downarrow}^{-} & -\Gamma_{\uparrow}%
^{+}-\Gamma_{\downarrow}^{+}%
\end{array}
\right]  \label{MatrixM}%
\end{equation}
depends on the total rates $\Gamma_{\sigma}^{\epsilon}=\sum_{j}\Gamma
_{j\sigma}^{\epsilon}$. This master equation treatment relies on a Markovian
approximation valid for frequencies $\omega$ lower than $\max[k_{B}%
T,\min\limits_{\sigma,i}(\left\vert E_{\sigma}-eV_{i}\right\vert )]/\hbar$
\cite{Engel}. From Eq. (\ref{MasterEquation}), the stationary occupation
probabilities $\bar{p}_{\psi}$ are
\begin{equation}
\bar{p}_{\sigma}=\frac{\Gamma_{\sigma}^{+}\Gamma_{-\sigma}^{-}+\Gamma
_{\sigma,-\sigma}(\Gamma_{\uparrow}^{+}+\Gamma_{\downarrow}^{+})}%
{\gamma_{\uparrow}\gamma_{\downarrow}-\Gamma_{\uparrow}^{+}\Gamma_{\downarrow
}^{+}+\sum\limits_{\sigma^{\prime}}(\gamma_{\sigma^{\prime}}+\Gamma
_{-\sigma^{\prime}}^{+})\Gamma_{\sigma^{\prime},-\sigma^{\prime}}}\;\text{,}
\label{p2}%
\end{equation}
with $\gamma_{\sigma}=\sum_{j}\gamma_{j\sigma}$, for $\sigma\in\left\{
\uparrow,\downarrow\right\}  $, and
\begin{equation}
\bar{p}_{0}=1-\bar{p}_{\uparrow}-\bar{p}_{\downarrow}\;\text{.} \label{p3}%
\end{equation}
These probabilities can be used to calculate the average value $\langle
I_{j}\rangle$ of the tunneling current $I_{j}(t)$ through junction $j$ as
$\langle I_{j}\rangle=\sum\limits_{\sigma}\langle I_{j,\sigma}\rangle$, where
$\langle I_{j,\sigma}\rangle=\sum\limits_{\epsilon}\langle I_{j,\sigma
}^{\epsilon}\rangle$ is the average current of electrons with spins $\sigma$,
and
\begin{equation}
\langle I_{j,\sigma}^{\epsilon}\rangle=e\epsilon\Gamma_{j\sigma}^{\epsilon
}\bar{p}_{A(\sigma,-\epsilon)}\text{ .} \label{curpart}%
\end{equation}
Here, $A(\sigma,\epsilon)$ is the state of the dot after the tunneling of an
electron with spin $\sigma$ in the direction $\epsilon$, i.e. $A(\sigma,-1)=0$
and $A(\sigma,+1)=\sigma$.

The frequency\ spectrum of the noise correlations can be defined as%

\begin{equation}
S_{ij}(\omega)=\int_{-\infty}^{+\infty}dt\mathcal{C}_{ij}(t)\exp(i\omega
t)\,\text{,} \label{CorrelationsDefinition}%
\end{equation}
where
\begin{equation}
\mathcal{C}_{ij}(t)=\langle\Delta I_{i}(t)\Delta I_{j}(0)\rangle+\langle\Delta
I_{i}(0)\Delta I_{j}(t)\rangle\text{ .} \label{CorrelationsDefinition2}%
\end{equation}
Following the method developed in Refs. \cite{setnoise}, we can write this
spectrum as:
\begin{equation}
S_{ij}(\omega)=\delta_{ij}S_{j}^{Sch}+S_{ij}^{c}(\omega) \label{eqnoise1}%
\end{equation}
with $S_{j}^{Sch}=\sum_{\sigma}S_{j\sigma}^{Sch}$ and $S_{ij}^{c}(\omega
)=\sum_{\sigma,\sigma^{\prime}}S_{i\sigma,j\sigma^{\prime}}^{c}(\omega)$.
Here,
\begin{equation}
S_{j\sigma}^{Sch}=2e\sum_{\epsilon}\left\vert \langle I_{j,\sigma}^{\epsilon
}\rangle\right\vert \text{ ,} \label{eqnoise2}%
\end{equation}
is the Schottky noise associated to the tunneling of electrons with spin
$\sigma$ through juntion $j$, and, from (\ref{MasterEquation}),
\begin{align}
\frac{S_{i\sigma,j\sigma^{\prime}}^{c}(\omega)}{2e^{2}}  &  =\sum
_{\epsilon,\epsilon^{\prime}}\epsilon\epsilon^{\prime}\left[  \Gamma_{i\sigma
}^{\epsilon^{\prime}}\widehat{\mathcal{G}}_{A(\sigma,-\epsilon^{\prime
}),A(\sigma^{\prime},\epsilon)}(\omega)\Gamma_{j\sigma^{\prime}}^{\epsilon
}\bar{p}_{A(\sigma^{\prime},-\epsilon)}\right. \label{eqnoise3}\\
&  \left.  +\Gamma_{{j}\sigma^{\prime}}^{\epsilon^{\prime}}\widehat
{\mathcal{G}}_{A(\sigma^{\prime},-\epsilon^{\prime}),A(\sigma,\epsilon
)}(-\omega)\Gamma_{i\sigma}^{\epsilon}\bar{p}_{A(\sigma,-\epsilon)}\right]
\text{ ,}\nonumber
\end{align}
with
\begin{equation}
\widehat{\mathcal{G}}(\omega)=-\operatorname{Re}\left[  \left(  i\omega
\widehat{I}+\widehat{M}\right)  ^{-1}\right]  \text{ } \label{G}%
\end{equation}
and $\widehat{I}$ the identity matrix. We also define, for later use, the spin
components of $S_{ij}(\omega)$:
\begin{equation}
S_{i\sigma,j\sigma^{\prime}}(\omega)=\delta_{ij}\delta_{\sigma\sigma^{\prime}%
}S_{i\sigma}^{Sch}+S_{i\sigma,j\sigma^{\prime}}^{c}(\omega)\;\text{.}
\label{SumSpectr}%
\end{equation}
Due to the existence of the stationary solution $\widehat{M}\bar{p}_{0}=0$,
the matrix $\widehat{M}$ has only two non-zero eigenvalues $\lambda_{+}$ and
$\lambda_{-}$, i.e. $\widehat{M}v_{\pm}=\lambda_{\pm}v_{\pm}$, given by%

\[
\lambda_{\pm}=\frac{1}{2}\left(  -\Lambda\pm\sqrt{\Lambda^{2}-4\Theta}\right)
<0\text{ ,}%
\]
with
\[
\Lambda=\Gamma_{\uparrow\downarrow}+\Gamma_{\downarrow\uparrow}+\gamma
_{\uparrow}+\gamma_{\downarrow}%
\]
and%

\[
\Theta=\gamma_{\uparrow}\gamma_{\downarrow}+\Gamma_{\uparrow\downarrow}\left(
\Gamma_{\downarrow}^{+}+\gamma_{\uparrow}\right)  +\Gamma_{\downarrow\uparrow
}\left(  \gamma_{\downarrow}+\Gamma_{\uparrow}^{+}\right)  -\Gamma_{\uparrow
}^{+}\Gamma_{\downarrow}^{+}\,\text{.}%
\]
Then, the matrix $\widehat{M}$ can be written in the form $\widehat
{M}=\widehat{R}^{-1}\left(  \lambda_{+}\widehat{E}_{+}+\lambda_{-}\widehat
{E}_{-}\right)  \widehat{R}$, where $\widehat{R}$ is a reversible $3\times3$
matrix, and $\widehat{E}_{+(-)}$ is a $3\times3$ matrix with the element $1$
at the first (second) row and first (second) column. Accordingly,
$\widehat{\mathcal{G}}(\omega)$ can be written as%

\begin{equation}
\widehat{\mathcal{G}}(\omega)=\frac{\widehat{\mathcal{A}}^{+}}{\omega
^{2}+\lambda_{+}^{2}}+\frac{\widehat{\mathcal{A}}^{-}}{\omega^{2}+\lambda
_{-}^{2}}\text{ ,} \label{Gg}%
\end{equation}
with $\widehat{\mathcal{A}}^{\pm}=-\lambda_{\pm}\widehat{R}^{-1}\widehat
{E}_{\pm}\widehat{R}$. Therefore, we have
\begin{equation}
S_{ij}^{c}(\omega)=\sum_{s=\pm}\frac{S_{ij}^{s}}{\omega^{2}+\lambda_{s}^{2}%
}\text{ ,} \label{resultSij}%
\end{equation}
where $S_{ij}^{s}$ follows from Eqs. (\ref{eqnoise3}) and (\ref{Gg}). The
total Schottky noise $S_{j}^{Sch}$ through junction $j$ is a white noise due
to the hypothesis of instantaneous tunneling. For a single junction biased by
a voltage source, one would get only this term. However, in the spectrum
$S_{ij}(\omega)$, interactions don't come into play only through the frequency
dependent term (\ref{resultSij}). Interactions also modify the values of the
terms $\langle I_{j\sigma}^{\epsilon}\rangle$ determining the Schottky noise.
Note that at high frequencies $\omega\gg\left\vert \lambda_{-}\right\vert $ ,
we have $S_{ij}\left(  \omega\right)  =\delta_{ij}S_{j}^{Sch}$. If we
furthermore assume $V\gg V_{\max}^{sgn(E_{0})}$, $S_{j}^{Sch}=2e\left\vert
\left\langle I_{j}\right\rangle \right\vert $ thus $S_{ij}\left(
\omega\right)  $ becomes Poissonian, i.e. $S_{ij}\left(  \omega\right)
=2e\left\vert \left\langle I_{i}\right\rangle \right\vert \delta_{ij}$.

In the three-terminal case studied here, we will be interested in the input
Fano factor
\[
F_{2}=\frac{S_{22}\left(  \omega=0\right)  }{2e\left\langle I_{2}\right\rangle
}\text{ ,}%
\]
and in the output cross-Fano factor
\[
F_{13}=\frac{S_{13}\left(  \omega=0\right)  }{2e\left\langle I_{2}%
\right\rangle }\text{ .}%
\]
We also define the resonance voltages
\[
V_{0}^{-}=\left\vert E_{0}\right\vert \frac{C}{eC_{2}}%
\]
and
\[
V_{0}^{+}=\left\vert E_{0}\right\vert \frac{C}{e\left(  C_{1}+C_{3}\right)
}\text{ .}%
\]
Since we consider $V>0$ only, at $B=0$, for $E_{0}$ positive (negative), the
dot orbital arrives at resonance with the Fermi level of the input (the
outputs) when $V\simeq V_{0}^{+(-)}$. If a magnetic field is applied, each of
these voltage resonances is split into two resonances
\[
V_{\uparrow(\downarrow)}^{-}=V_{0}^{-}+(-)\frac{g\mu_{B}BC}{2eC_{2}}%
\]
and
\[
V_{\uparrow(\downarrow)}^{+}=V_{0}^{+}-(+)\frac{g\mu_{B}BC}{2e\left(
C_{1}+C_{3}\right)  }\text{ ,}%
\]
associated to the $\uparrow(\downarrow)$ levels respectively because we
consider $B>0$ only. We expect $F_{2}$ and $F_{13}$ to show strong variations
for $V\simeq V_{\uparrow(\downarrow)}^{sgn(E_{0})}$.

\subsection{C. Time-domain analysis}

The correlation function $\mathcal{C}_{ij}(t)$ can be obtained from the
inverse Fourier transform of Eqs. (\ref{eqnoise1}), (\ref{eqnoise2}) and
(\ref{resultSij}):%

\begin{equation}
\mathcal{C}_{ij}(t)=\delta_{ij}\delta(t)S_{j}^{Sch}+\sum_{s=\pm}\frac
{S_{ij}^{s}}{2\left|  \lambda_{s}\right|  }\exp(-\left|  t\right|  \left|
\lambda_{s}\right|  )\text{ .} \label{correlationC13}%
\end{equation}
In the sequential tunneling limit, tunneling events occur one by one, thus%

\begin{equation}
\lim_{t\rightarrow0^{+}}\mathcal{C}_{ij}(t)=-2\left\langle I_{i}\right\rangle
\left\langle I_{j}\right\rangle <0\text{ .} \label{Cprop0}%
\end{equation}
Let us first focus on the spin-degenerate case, that is $\Gamma_{j\uparrow
}^{\epsilon}=\Gamma_{j\downarrow}^{\epsilon}$ for $j\in\left\{  1,2,3\right\}
$. In this case, the eigenvectors $v_{+/-}$ of $\widehat{M}$ correspond to the
spin/charge excitations of the system (i.e. $v_{+}\sim\left[  1,-1,0\right]
$, $v_{-}\sim\left[  1,1,-2\right]  $), and $\lambda_{+/-}$ to their
relaxation rates. This is directly connected to the fact that in the
spin-degenerate case, $S_{ij}^{+}=0$, thus $S_{ij}(\omega)-\delta_{ij}%
S_{j}^{Sch}$ is a Lorentzian function and $\mathcal{C}_{ij}(t)-\delta
_{ij}\delta(t)S_{j}^{Sch}=S_{ij}^{-}\exp(-\left\vert t\right\vert \left\vert
\lambda_{-}\right\vert )/2\left\vert \lambda_{-}\right\vert .$ This last
equation implies that, for any time, $\mathcal{C}_{22}(t)-\delta(t)S_{2}%
^{Sch}$ and $\mathcal{C}_{13}(t)$ keep the same sign, which is negative
according to Eq. (\ref{Cprop0}). Thus, \textit{in the spin-degenerate
one-orbital case}, $F_{2}$ \textit{is always sub-Poissonian and} $F_{13}$
\textit{always negative}. When spin-degeneracy is lifted, $v_{+/-}$ both
become a linear combination of the charge and spin excitations. Thus, having
$S_{ij}^{+}\neq0$ is not forbidden anymore. Eqs. (\ref{correlationC13}), and
(\ref{Cprop0}) altogether with $\left\vert \lambda_{+}\right\vert <\left\vert
\lambda_{-}\right\vert $ imply that if $S_{ij}(\omega=0)>\delta_{ij}%
S_{j}^{Sch}$, one has $S_{ij}^{-}<0$ and $S_{ij}^{+}>0$. Therefore, in the
one-orbital case, a positive sign for $F_{2}-S_{2}^{Sch}/2e\left\langle
I_{2}\right\rangle $ and $F_{13}$\ can only be due to terms in $\lambda_{+}$.

The results obtained for $\mathcal{C}_{ij}(t)$ can be put in perspective with
some fundamental quantities like the average dwell time $t_{\sigma}$ of spins
$\sigma$ on the dot and the average delay $t_{0}$ between the occupancy of the
dot by two consecutive electrons. These quantities can be calculated for
$\gamma_{sf}=0$ as
\begin{equation}
t_{\sigma}=\frac{4e^{2}\bar{p}_{\sigma}}{%
%TCIMACRO{\dsum \limits_{j}}%
%BeginExpansion
{\displaystyle\sum\limits_{j}}
%EndExpansion
S_{j,\sigma}^{Sch}}\text{ } \label{a}%
\end{equation}
and
\begin{equation}
t_{0}=\frac{4e^{2}\bar{p}_{0}}{%
%TCIMACRO{\dsum \limits_{j}}%
%BeginExpansion
{\displaystyle\sum\limits_{j}}
%EndExpansion
S_{j}^{Sch}}\text{ .} \label{b}%
\end{equation}
The noise reaches its high-voltage limit once $V\gg V_{\max}^{sgn(E_{0})}%
=\max_{\sigma}(V_{\sigma}^{sgn(E_{0}})$ with $V_{\max}^{+}=V_{\downarrow}^{+}$
and $V_{\max}^{-}=V_{\uparrow}^{-}$. In this limit, the current transport is
unidirectional, i.e. $\left\langle I_{2,\sigma}^{-}\right\rangle =0$ and
$\left\langle I_{j,\sigma}^{+}\right\rangle =0$ and for any $j\times\sigma
\in\left\{  1,3\right\}  \times\left\{  \uparrow,\downarrow\right\}  $. Thus,
Eqs. (\ref{a}) and (\ref{b}) lead to $t_{0}=1/2\gamma_{2}$ and $t_{\sigma
}=1/\left(  \gamma_{1\sigma}+\gamma_{3\sigma}\right)  $. The average number
\begin{equation}
n_{b}=\frac{S_{2\uparrow}^{Sch}}{S_{2\downarrow}^{Sch}} \label{nb}%
\end{equation}
of up spins crossing the input junction\ between two consecutive down spins
for $\gamma_{sf}=0$, which becomes $n_{b}=I_{2\uparrow}/I_{2\downarrow}$ for
$V\gg V_{\max}^{sgn(E_{0})}$, is also of importance. It can be used to
calculate the average duration
\begin{equation}
t_{b}=n_{b}t_{\uparrow}+\left(  n_{b}+1\right)  t_{0} \label{tb}%
\end{equation}
between the occupation of the dot by two consecutive down spins for
$\gamma_{sf}=0$. In Section III.C, the analysis of $\mathcal{C}_{ij}(t)$ will
be supplemented by simulating numerically the time evolution of the spin
$\sigma_{dot}$ of the dot. As expected, these simulations are in agreement
with the results obtained from the master equation approach, but their
interest is to allow a visualization of $\sigma_{dot}(t)$.

\subsection{D. Relation between $F_{2}$ and $F_{13}$}

The average input current $\left\langle I_{2}\right\rangle $ and the input
Fano factor $F_{2}$ in a three-terminal device correspond to the average
current and the Fano factor in a two-terminal device where the output leads
$1$ and $3$ are replaced by an effective output with a net spin-independent
tunneling rate $\gamma_{t}=\gamma_{1}+\gamma_{3}$ and with an effective
polarization $P_{out}=\left(  \gamma_{1}P_{1}+\gamma_{3}P_{3}\right)
/\gamma_{t}$. Then, one fundamental question to answer is whether there is a
simple relation between $F_{2}$ and $F_{13}$ in the three-terminal circuit.
Charge conservation and the finite dispersion of $\left\vert \sigma
_{dot}(t)\right\vert $ lead to \cite{setnoise}
\begin{equation}
S_{22}\left(  \omega=0\right)  =S_{11}\left(  \omega=0\right)  +S_{33}\left(
\omega=0\right)  +2S_{13}\left(  \omega=0\right)  \text{ .} \label{sum2}%
\end{equation}
At high voltages $V\gg V_{\max}^{sgn(E_{0})}$, the unidirectionality of
current transport and the average-currents conservation lead to $S_{2}%
^{Sch}=S_{1}^{Sch}+S_{3}^{Sch}$. In this limit, Eqs. (\ref{eqnoise1}) and
(\ref{sum2}) imply that
\begin{equation}
S_{22}^{c}\left(  \omega=0\right)  =S_{11}^{c}\left(  \omega=0\right)
+S_{33}^{c}\left(  \omega=0\right)  +2S_{13}^{c}\left(  \omega=0\right)
\text{ .} \label{sumSc}%
\end{equation}
Since the voltage bias is the same for leads $1$ and $3$, we have
$\Gamma_{3,\sigma}^{\epsilon}/\Gamma_{1,\sigma}^{\epsilon}=\gamma_{1\sigma
}/\gamma_{3\sigma}$ for $\epsilon=\pm1$. Then, from (\ref{eqnoise3}), in our
singly-occupied one-orbital case, Eq. (\ref{sumSc}) leads to
\begin{multline*}
S_{22}^{c}\left(  \omega=0\right)  =\\
\sum_{\sigma,\sigma^{\prime}}\frac{\left(  \gamma_{1\sigma}+\gamma_{3\sigma
}\right)  \left(  \gamma_{1\sigma^{\prime}}+\gamma_{3\sigma^{\prime}}\right)
}{\gamma_{1\sigma}\gamma_{3\sigma^{\prime}}}S_{1,\sigma,3,\sigma^{\prime}}%
^{c}\left(  \omega=0\right)  \text{ .}%
\end{multline*}
If we furthermore assume $P_{1}=P_{3}$, the ratio $\gamma_{1\sigma}%
/\gamma_{3\sigma}=\gamma_{1}/\gamma_{3}$ is independent of $\sigma$ and
\begin{equation}
F_{13}=\left(  F_{2}-1\right)  \frac{\gamma_{1}\gamma_{3}}{\gamma_{t}^{2}%
}\text{ .} \label{SFrelation}%
\end{equation}
In summary, for the one-orbital circuit studied here, there exists a simple
relation between $F_{2}$ and $F_{13}$ when $P_{1}=P_{3}$ and $V\gg V_{\max
}^{sgn(E_{0})}$. Note that the derivation of property (\ref{SFrelation})
requires neither $\gamma_{sf}=0$, nor $B=0$. On the contrary, the voltage-bias
configuration used is crucial. Indeed, if the three leads $1$, $2$, $3$, were
for instance biased at voltages $V$, $V/2$ and $0$ respectively, the current
transport would not be unidirectional even in the high $V$-limit. When
property (\ref{SFrelation}) is verified, a super-Poissonian (sub-Poissonian)
$F_{2}$ is automatically associated with positive (negative) zero-frequency
cross-correlations. However, Sections III and IV, which also treat this
one-orbital case, illustrate that when $P_{1}\neq P_{3}$ or $V\lesssim
V_{\max}^{sgn(E_{0})}$, property (\ref{SFrelation}) is not valid anymore, and
in particular a super-Poissonian $F_{2}$ can be obtained without a positive
$F_{13}$. In Section IV.B, $F_{2}$ and $F_{13}$ even show variations which are
\textit{qualitatively} different: $F_{13}$ displays a voltage resonance not
present in $F_{2}$. Thus, even for the one-orbital quantum dot circuit studied
here, the three-terminal problem is in general not trivially connected to the
two-terminal problem.

The main ingredients for deriving (\ref{SFrelation}) are the unidirectionality
of current transport and a division of current between the two outputs
identical for the two spin directions. One can wonder whether any
tunnel-junction circuit with a geometry analoguous to that of Figure
\ref{Pict1} satisfies property (\ref{SFrelation}) for $V\gg V_{\max
}^{sgn(E_{0})}$ and $P_{1}=P_{3}$. Indeed, it is sometimes the case. For
instance, B\"{o}rlin \textit{et al.} have studied at $T=0$ a normal-metal
island too large to have charging effects, connected, through tunnel
junctions, to one superconducting or normal input lead and to two
normal\ output leads with $P_{1}=P_{3}=0$ placed at the same output potential
\cite{p6}. For this system, in both the hybrid and the normal cases, a
relation analog to (\ref{SFrelation}) is fulfilled, provided $\gamma
_{1}/\gamma_{3}$ is replaced by $g_{1}/g_{3}$, where $g_{1}$ and $g_{3}$ are
the conductances of the output junctions. In spite of this, (\ref{SFrelation})
is not universal even for \textit{spin-degenerate} tunnel-junction circuits.
This can be shown by considering the circuit of Fig. \ref{Pict1}, with $B=0$,
$P_{1}=P_{2}=P_{3}=0$ and a \textit{two-orbital} dot (Section V). In this
case, the division of currents between the two outputs will generally depend
on the orbital considered, because of the different spatial extensions of the
orbitals and of the asymmetric positions of the output leads with respect to
them \cite{Leturcq}. One has to assume that the division of currents between
leads 1 and 3 is independent of the orbital considered in order to recover
property (\ref{SFrelation}) at $V\gg V_{\max}^{sgn(E_{0})}$.

\subsection{E. Influence of screening currents at non-zero frequencies}

The total instantaneous current $I_{j}^{tot}(t)$ passing through branch $j$
includes the tunneling current $I_{j}(t)$ but also the screening currents
needed to guarantee the electrostatic equilibrium of the capacitors after\ a
tunneling event through any junction $i\in\left\{  1,2,3\right\}  $. However,
screening currents contribute neither to the average value $\left\langle
I_{j}^{tot}\right\rangle $ of the total current $I_{j}^{tot}(t)$, i.e.
$\left\langle I_{j}^{tot}\right\rangle =\left\langle I_{j}\right\rangle $, nor
to the low frequency part of the total current correlations $S_{ij}%
^{tot}(\omega)$, i.e. $S_{ij}^{tot}(\omega)=S_{ij}(\omega=0)$ for $\left\vert
\omega\right\vert \ll\left\vert \lambda_{+}\right\vert $, because, in average,
the screening currents due to tunneling through the different junctions
compensate each other at zero frequency (see for instance \cite{blanter:00}).
Screening currents contribute to $S_{ij}^{tot}(\omega)$ only once
$S_{ij}(\omega)$ deviates from its zero-frequency limit\textit{. }In the
following, we will assume that the cutoff frequency $\left\vert \lambda
_{+}\right\vert $ is much larger than the inverse of the collective response
times associated to the charging of the capacitors. This is equivalent to
assuming that, on the time scale on which $\sigma_{dot}(t)$ varies, any charge
variation of the dot triggers\ instantaneously the screening currents needed
for its compensation:
\begin{equation}
I_{j}^{tot}(t)=I_{j}(t)-\frac{C_{j}}{C}\sum\limits_{i}I_{i}(t)\text{ .}
\label{screeningC}%
\end{equation}
According to this approximation, the total current correlations $S_{ij}%
^{tot}(\omega)$, including screening currents, can be written as
\begin{equation}
S_{ij}^{tot}(\omega)=\sum_{n,m}\left(  \delta_{in}-\frac{C_{i}}{C}\right)
\left(  \delta_{jm}-\frac{C_{j}}{C}\right)  S_{nm}(\omega)\text{ .}
\label{Sijtot}%
\end{equation}
The sign of these total current cross-correlations is not trivial. This
problem is addressed in Section III.E.

\section{III. One-orbital quantum dot connected to ferromagnetic leads, in the
absence of a magnetic field}

Here, we focus on the one-orbital case introduced in Section II, for $B=0$ and
magnetically polarized leads. In the absence of a magnetic field, one single
resonance is expected in the voltage characteristics, for $V\simeq
V_{0}^{sgn(E_{0})}$. Figures \ref{Pict2} to \ref{Pict6} show curves for a
constant value of the polarization amplitudes $\left|  P_{1}\right|  =\left|
P_{2}\right|  =\left|  P_{3}\right|  =0.6$. This corresponds for instance to
having the different leads made out of the same ferromagnetic material.

\subsection{A. Zero-frequency results for $E_{0}>0$}

\begin{figure}[h]
\includegraphics[width=1.\linewidth]{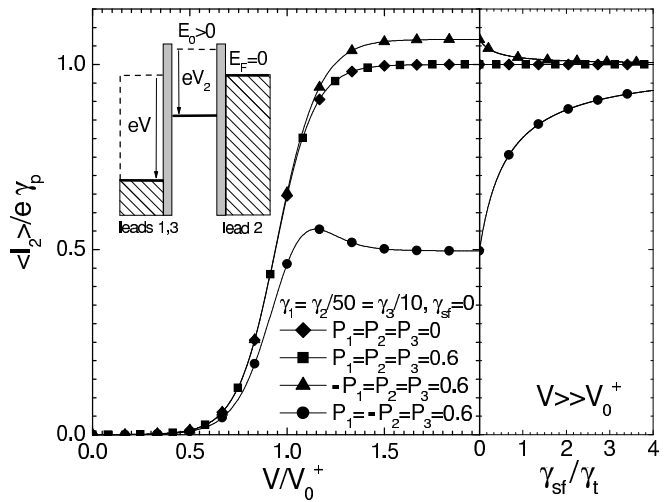}\caption{Left panel:
Current-voltage characteristic of the circuit of Fig. \ref{Pict1}, for
$E_{0}>0$, $C_{1}=C_{2}=C_{3}$, $\gamma_{1}=\gamma_{2}/50=\gamma_{3}/10$,
$k_{B}T/\left\vert E_{0}\right\vert =0.1$, $B=0$, and different values of lead
polarizations. The average current $\left\langle I_{2}\right\rangle $ through
lead $2$ is plotted in units of its paramagnetic high voltage limit
$e\gamma_{p}=2e\gamma_{2}(\gamma_{1}+\gamma_{3})/(\gamma_{1}+2\gamma
_{2}+\gamma_{3})$; the voltage in units of the resonance voltage $V_{0}^{+}$
(see II.B). For $P_{1}=P_{2}=P_{3}$ (squares), $\left\langle I_{2}%
\right\rangle $ coincides with the paramagnetic case (diamonds). In the other
cases, the high-voltage limit of $\left\langle I_{2}\right\rangle $ can be
larger or smaller than the paramagnetic value, depending on the lead
polarizations. The inset shows the electrochemical potentials in the circuit.
Notation\textbf{\ }$E_{F}$ refers to the Fermi level in lead 2. (In all the
plots, potentials are shown for the case where the dot is empty\textbf{).
}Right panel: Influence of spin-flip scattering in the high-voltage limit
$V\gg V_{0}^{+}$. Here, the spin flip scatterring rate $\gamma_{sf}$ is
expressed in units of $\gamma_{t}=\gamma_{1}+\gamma_{3}$. Spin-flip scattering
makes the $\left\langle I_{2}\right\rangle (V)$ curve tend to the paramagnetic
one.}%
\label{Pict2}%
\end{figure}\begin{figure}[hh]
\includegraphics[width=1.\linewidth]{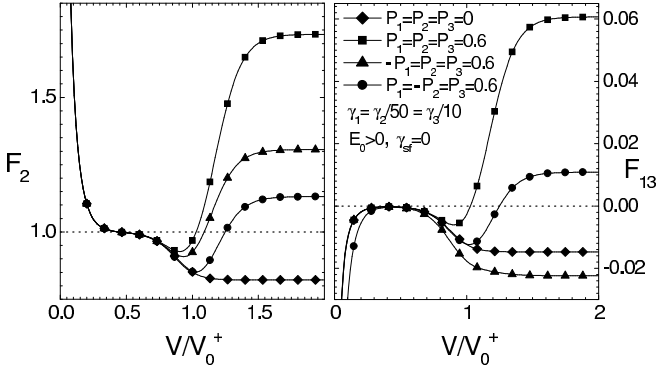}\caption{Input Fano factor
$F_{2}=S_{22}(\omega=0)/2eI_{2}$ (left panel) and output cross-Fano factor
$F_{13}=S_{13}(\omega=0)/2eI_{2}$ between leads $1$ and $3$ (right panel) as a
function of the bias voltage $V$ for $\gamma_{sf}=0$. The curves are shown for
the same circuit parameters as in Fig.~\ref{Pict2}. When $P_{1}=P_{2}=P_{3}$
(squares), $F_{2}$ is different from that of the paramagnetic case (diamonds)
in contrast to what happens for $\left\langle I_{2}\right\rangle $. At high
enough voltages, the cross-correlations are positive in the cases
$P_{1}=-P_{2}=P_{3}=0.6$ (circles) and $P_{1}=P_{2}=P_{3}=0.6$. Note that the
sign of the cross-correlations can be reversed by changing the sign of $P_{1}
$. The case $-P_{1}=P_{2}=P_{3}=0.6$ (triangles) illustrates that having a
super-Poissonian $F_{2}$ is not sufficient to obtain a positive $F_{13}$.}%
\label{Pict3}%
\end{figure}\begin{figure}[hhh]
\includegraphics[width=1.\linewidth]{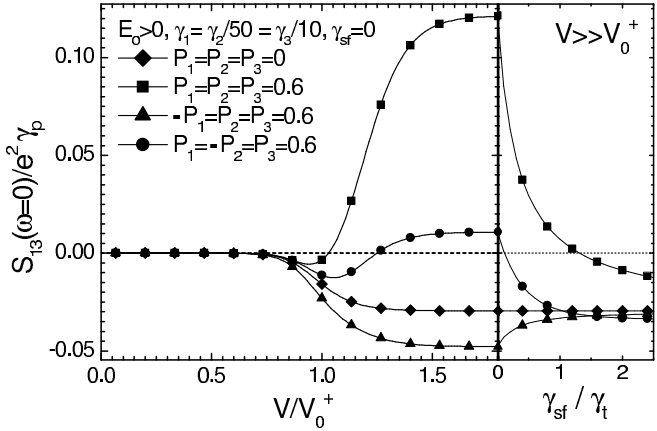}\caption{Zero-frequency
cross-correlations $S_{13}(\omega=0)$ between leads $1$ and $3$ as a function
of the bias voltage $V$ for $\gamma_{sf}=0$ (left panel) and as a function of
$\gamma_{sf}$ for $V\gg V_{0}^{+}$ (right panel). The curves are shown for the
same circuit parameters as in Fig.~\ref{Pict2}. In the paramagnetic case
(diamonds), spin-flip scattering has no effect. In the limit of large
$\gamma_{sf}$, cross-correlations tend to the paramagnetic value.}%
\label{Pict4}%
\end{figure}

We first consider the case in which the orbital level $E_{0}$ is above the
Fermi level of the leads at equilibrium ($E_{0}>0$). The typical voltage
dependence of the average input current $\langle I_{2}\rangle$ is shown in the
left panel of Fig.~\ref{Pict2}. This current is exponentially suppressed at
low voltages, increases around the voltage $V_{0}^{+}$ and saturates at higher
voltages. The width on which $\langle I_{2}\rangle$ varies is of the order of
$\Delta V\sim10k_{B}TC/e\left(  C_{1}+C_{3}\right)  $. The high-voltage limit
of $\left\langle I_{2}\right\rangle $ depends on the polarizations $P_{i}$ and
rates $\gamma_{i}$ but not on the capacitances $C_{i}$ because the tunneling
rates saturate at high voltages [see Eq. (\ref{rates})]. For the paramagnetic
case, this limit is
\begin{equation}
e\gamma_{p}=e\frac{2\gamma_{2}\gamma_{t}}{\gamma_{t}+2\gamma_{2}}\text{ .}
\label{curPara}%
\end{equation}
In this last expression, $\gamma_{2}$ is weighted by a factor 2 to account for
both the populations of up and down spins arriving from the input lead. The
rate $\gamma_{t}=\gamma_{1}+\gamma_{3}$ is not weighted by such a factor
because there can be only one electron at a time on the dot, which can tunnel
to the output leads with the total rate $\gamma_{t}$. For a sample with
magnetic contacts, the high-voltage limit of $\langle I_{2}\rangle$ can be
higher or lower than $e\gamma_{p}$, depending on the parameters considered.
Indeed, for $V\gg V_{0}^{+}$, we have $I_{2}(P_{1},P_{2},P_{3})-I_{2}%
(0,0,0)=e\gamma_{p}P_{out}\langle\sigma_{dot}\rangle$, where $P_{out}%
=(P_{1}\gamma_{1}+P_{3}\gamma_{3})/\gamma_{t}$ is the net output lead
polarization, and where $\langle\sigma_{dot}\rangle=\nu(P_{2}-P_{out})$ is the
average spin of the dot. Here, $\nu$ is a positive function of the
polarizations, tunneling and scattering rates, which tends to $0$ at large
$\gamma_{sf}$. For $P_{1}=P_{2}=P_{3}$, the current is the same as in the
paramagnetic case because the populations of spin are matched between the
input and the output, thus $\langle\sigma_{dot}\rangle=0$. Having a saturation
current different from the paramagnetic case requires $P_{out}\neq0$
\textit{and} $\langle\sigma_{dot}\rangle\neq0$. When $\left\vert
P_{out}\right\vert >\left\vert P_{2}\right\vert $, the high-voltage limit of
$\left\langle I_{2}\right\rangle $ is lower than that of the paramagnetic case
because the spins in minority at the output block the dot, which leads to a
$\langle\sigma_{dot}\rangle$ with the same sign as $-P_{out}$. In this case,
$\left\langle I_{2}\right\rangle $ can show negative differential resistance
above $V_{0}^{+}$, due to the deblockade of the dot by thermal fluctuations
which can send back the blocking spins to electrode 2 for $V\simeq V_{0}^{+}$
\cite{bulka:99}. Spin-flip scattering modifies the $\left\langle
I_{2}\right\rangle (V)$ curve once $\gamma_{sf}$ is of the order of the
tunneling rates. It suppresses spin accumulation and makes the $\left\langle
I_{2}\right\rangle (V)$ curve tend to the paramagnetic one.

Figure~\ref{Pict3} shows $F_{2}$ and $F_{13}$ as a function of $V$ for
$\gamma_{sf}=0$. We also show in Fig. \ref{Pict4} the zero-frequency
cross-correlations $S_{13}\left(  \omega=0\right)  $ because it is the signal
measured in practice. Well below $V_{0}^{+}$, $S_{13}\left(  \omega=0\right)
$ is exponentially suppressed like $\left\langle I_{2}\right\rangle $ because
there are very few tunneling events. In this regime, the dot is empty most of
the time, and when an electron arrives on the dot, it leaves it with a much
higher rate ($\Gamma_{\sigma}^{-}\gg\Gamma_{\sigma}^{+}$): the electronic
transport is limited only by thermally activated tunneling through junction 2.
Tunneling events are thus uncorrelated and $F_{2}$ is Poissonian, with a
unitary plateau following the thermal divergence $2k_{B}T/eV$\ occurring at
$V=0$. For the same reasons, $F_{13}$ displays a zero plateau after a
polarization-dependent thermal peak at $V=0$. Around $V\simeq V_{0}^{+}$,
$F_{2}$, $F_{13}$ and $S_{13}\left(  \omega=0\right)  $ strongly vary. The
high-voltage limit depends on tunneling rates and polarizations. In the
paramagnetic case, the high-voltage limit of $F_{2}$ lies in the interval
$[1/2,1]$, and that of $F_{13}$ in $[-1/8,0]$. In the ferromagnetic case, the
high-voltage limit of $F_{2}$ can be either sub- or super-Poissonian, as
already pointed out in the two-terminal case \cite{bulka:99}. Spin
accumulation is not a necessary condition for having a super-Poissonian
$F_{2}$, as can be seen for $P_{1}=P_{2}=P_{3}$, where $\left\langle
\sigma_{dot}\right\rangle =0.$ Negative differential resistance is not
necessary either (see case $P_{1}=P_{2}=P_{3}=0.6$ in Figs. \ref{Pict2} and
\ref{Pict3}). Cross-correlations can be either positive or negative depending
on the parameters considered, as shown by Figs. \ref{Pict3} and \ref{Pict4}.
Interestingly, the sign of cross-correlations can be switched by reversing the
magnetization of one contact. The case $P_{1}=P_{2}=P_{3}=0.6$ of
Figs.~\ref{Pict3}, \ref{Pict4} corresponds to a super-Poissonian $F_{2}$
\textit{and} a positive $F_{13}$. The case $-P_{1}=P_{2}=P_{3}=0.6$ shows that
a super-Poissonian $F_{2}$ is not automatically associated with positive
output cross-correlations. In this case, the cross-correlations are even more
negative than in the paramagnetic case. This will be explained\ physically in
Section III.C.

The effect of spin-flip scattering on $S_{13}(\omega=0)$ is shown in the right
panel of Fig. \ref{Pict4}. In the paramagnetic case, spin-flip scattering has
no effect on $S_{13}(\omega=0)$. In the ferromagnetic case, when $\gamma_{sf}$
\ is of the order of the tunneling rates, $S_{13}(\omega=0)$ is modified. In
the high-$\gamma_{sf}$ limit, cross-correlations tend to the paramagnetic case
for any value of the polarizations. Thus, strong spin-flip scattering
suppresses positive cross-correlations. However, in practice, it is possible
to make quantum dots connected to ferromagnetic leads with spin-flip rates
much smaller than the tunneling rates \cite{Deshmukh}. Hence, spin-flip
scattering should not be an obstacle for observing positive cross-correlations
in the quantum-dot circuit studied here.

\subsection{B. Zero-frequency results for $E_{0}<0$}

\begin{figure}[h]
\includegraphics[width=0.9\linewidth]{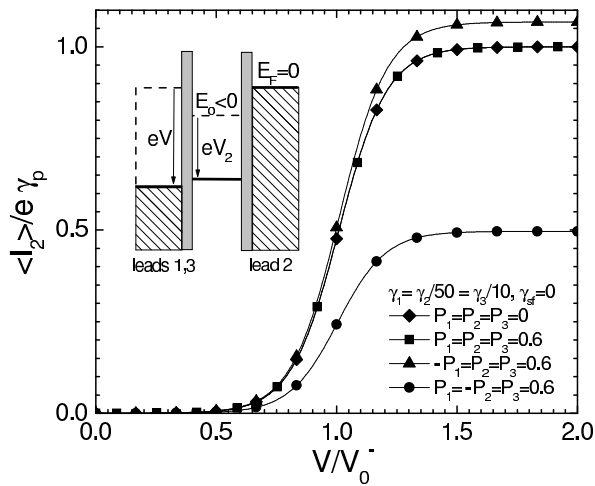}\caption{Current-voltage
characteristic of the circuit of Fig. \ref{Pict1} for $E_{0}<0$. The
polarizations, tunnel rates, capacitances and reduced temperature
$k_{B}T/\left\vert E_{0}\right\vert $ used are the same as in Fig.
\ref{Pict2}, plotted for $E_{0}>0$. The results differ from the case $E_{0}>0$
only for $V\simeq V_{0}^{-}$.}%
\label{Pict5}%
\end{figure}\begin{figure}[hh]
\includegraphics[width=1.\linewidth]{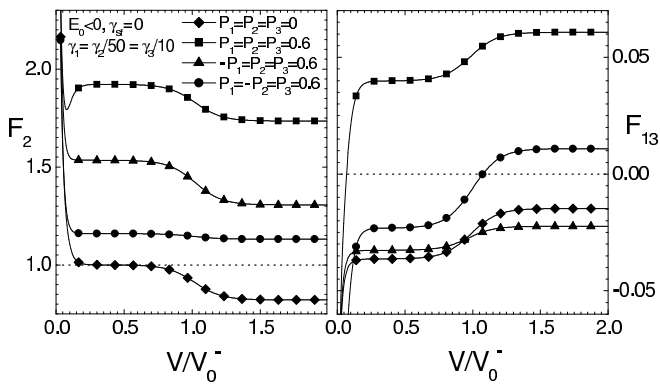}\caption{Input Fano factor
$F_{2}$ and cross-Fano factor $F_{13}$ as a function of the bias voltage $V$.
The curves are shown for the same circuit parameters as in Fig.~\ref{Pict5}.}%
\label{Pict6b}%
\end{figure}

\begin{figure}[h]
\includegraphics[width=0.9\linewidth]{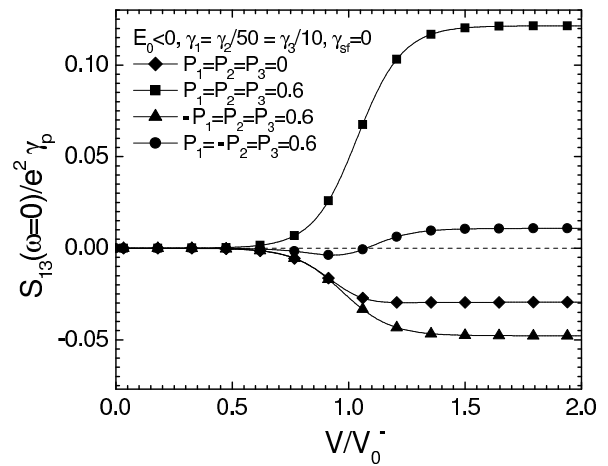}\caption{Zero-frequency
current cross-correlations $S_{13}(\omega=0) $ between leads $1$ and $3$ as a
function of the bias voltage $V$. The curves are shown for the same circuit
parameters as in Fig.~\ref{Pict5}.}%
\label{Pict6}%
\end{figure}

We now discuss the case in which the orbital level $E_{0}$ is below the Fermi
level of the leads at equilibrium ($E_{0}<0$). First, in the low voltage limit
in which very few electrons can flow through the device, $\left\langle
I_{2}\right\rangle $ and $S_{13}\left(  \omega=0\right)  $ exponentially tend
to zero like in the case $E_{0}>0$ (Figs. \ref{Pict5} and \ref{Pict6}).
However, for $F_{2}$ and $F_{13}$, the results differ (Fig. \ref{Pict6b}).
Above the $2k_{B}T/eV$\ thermal peak, the low voltage plateau of $F_{2}$ is
always super-Poissonian for $P_{out}\neq0$. Above a polarization-dependent
thermal peak, $F_{13}$ displays a low voltage plateau which is either negative
or positive. This features indicate a correlated transfer of charges in spite
of the thermally activated limit. In fact, for $V\ll V_{0}^{-}$, the dot is
occupied most of the time and the electronic transport is limited by thermally
activated tunneling through the output junctions 1 and 3. In these conditions,
contrarily to what happens for $E_{0}>0$, the polarization of the output leads
comes into play even for $V\longrightarrow0$. Indeed, when $P_{out}\neq0$, the
spins in minority at the output have less chances to leave the dot under the
effect of thermal fluctuations. In the intermediate voltage range $V\simeq
V_{0}^{-}$, the quantities $\left\langle I_{2}\right\rangle $, $F_{2}$,
$F_{13}$ and $S_{13}(\omega=0)$ differ from the case $E_{0}>0$. However, at
$V\gg V_{0}^{-}$, they take the same values as for $E_{0}>0$ and $V\gg
V_{0}^{+}$. In this high-voltage limit, the effect of spin flip scattering is
identical to that of the case $E_{0}>0$. In particular, the right panels of
Figs. \ref{Pict2} and \ref{Pict4} are also valid for $E_{0}<0$.

\subsection{C. Interpretation of these zero-frequency results: Dynamical
Spin-Blockade}

\begin{table}[h]
$
\begin{array}
[c]{c}%
\begin{tabular}
[c]{|c|c|c|c|c|c|c|c|c|}\hline
case & $\frac{S_{13}}{e^{2}\gamma_{p}}$ & $\frac{S_{1\uparrow,3\uparrow}%
}{e^{2}\gamma_{p}}$ & $\frac{S_{1\downarrow,3\downarrow}}{e^{2}\gamma_{p}}$ &
$\frac{S_{1\uparrow,3\downarrow}}{e^{2}\gamma_{p}}$ & $\frac{S_{1\downarrow
,3\uparrow}}{e^{2}\gamma_{p}}$ & $\frac{I_{1\uparrow}}{I_{3\uparrow}}$ &
$\frac{I_{1\downarrow}}{I_{3\downarrow}}$ & $n_{b}$\\\hline
$\blacklozenge$ & -0.030 & -0.007 & -0.007 & -0.007 & -0.007 & 0.1 & 0.1 &
1\\\hline
$\blacksquare$ & 0.121 & 0.149 & -0.013 & -0.007 & -0.007 & 0.1 & 0.1 &
4\\\hline
$\blacktriangle$ & -0.048 & 0.026 & -0.029 & -0.003 & -0.042 & 0.025 & 0.4 &
4\\\hline
$\bullet$ & 0.011 & 0.005 & -0.011 & 0.008 & 0.008 & 0.1 & 0.1 & 0.25\\\hline
\end{tabular}
\\
\vspace*{0.1cm}\\%
\begin{tabular}
[c]{|c|c|c|c|c|c|c|c|c|c|}\hline
case & $p_{\uparrow}$ & $p_{\downarrow}$ & $p_{0}$ & $\gamma_{p}t_{\uparrow}$
& $\gamma_{p}t_{\downarrow}$ & $\gamma_{p}t_{0}$ & $t_{b}\gamma_{p}$ &
$\frac{\gamma_{p}}{\left\vert \lambda^{-}\right\vert }$ & $\frac{\gamma_{p}%
}{\left\vert \lambda^{+}\right\vert }$\\\hline
$\blacklozenge$ & 0.450 & 0.450 & 0.100 & 0.90 & 0.90 & 0.10 & 1.10 & 0.09 &
0.90\\\hline
$\blacksquare$ & 0.450 & 0.450 & 0.099 & 0.56 & 2.25 & 0.10 & 2.75 & 0.09 &
1.46\\\hline
$\blacktriangle$ & 0.516 & 0.378 & 0.106 & 0.60 & 1.77 & 0.10 & 2.91 & 0.09 &
1.31\\\hline
$\bullet$ & 0.056 & 0.895 & 0.049 & 0.563 & 2.25 & 0.10 & 0.26 & 0.09 &
0.68\\\hline
\end{tabular}
\end{array}
$\caption{Top: \textit{Zero-frequency} output cross-correlations
$S_{13}(\omega=0)$ and its spin contributions $S_{1\sigma,3\sigma^{\prime}%
}(\omega=0)$, division $I_{1\sigma}/I_{3\sigma}$ of spin currents between
leads $1$ and $3$ and average number $n_{b}$ of up spins crossing the input
junction\ between two consecutive down spins, for the high-voltage limit $V\gg
V_{0}^{sng(E_{0})}$ of the cases studied in Sections III.A and III.B (Figs.
\ref{Pict2} to \ref{Pict6}). Bottom: Probabilities $p_{\psi}$ and comparison
of the different timescales of the system, for the same parameters (The
summation rules (\ref{p3}) and (\ref{SumSpectr}) are not exactly verified by
the values given in this table because of the limitation in the number of
digits).}%
\label{TableI}%
\end{table}\begin{figure}[hh]
\includegraphics[width=1.\linewidth]{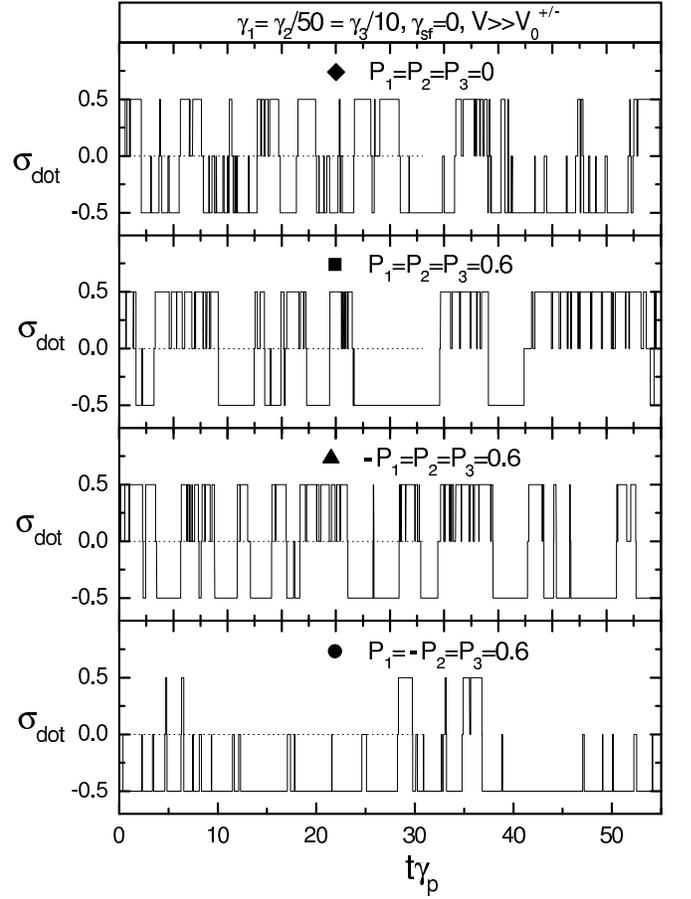}\caption{Numerical simulation
of the spin $\sigma_{dot}$ of the dot as a function of time in the limit $V\gg
V_{0}^{sng(E_{0})}$, for the different cases considered in Figs. \ref{Pict2}
to \ref{Pict6}.}%
\label{Pict7}%
\end{figure}\begin{figure}[hhh]
\includegraphics[width=0.9\linewidth]{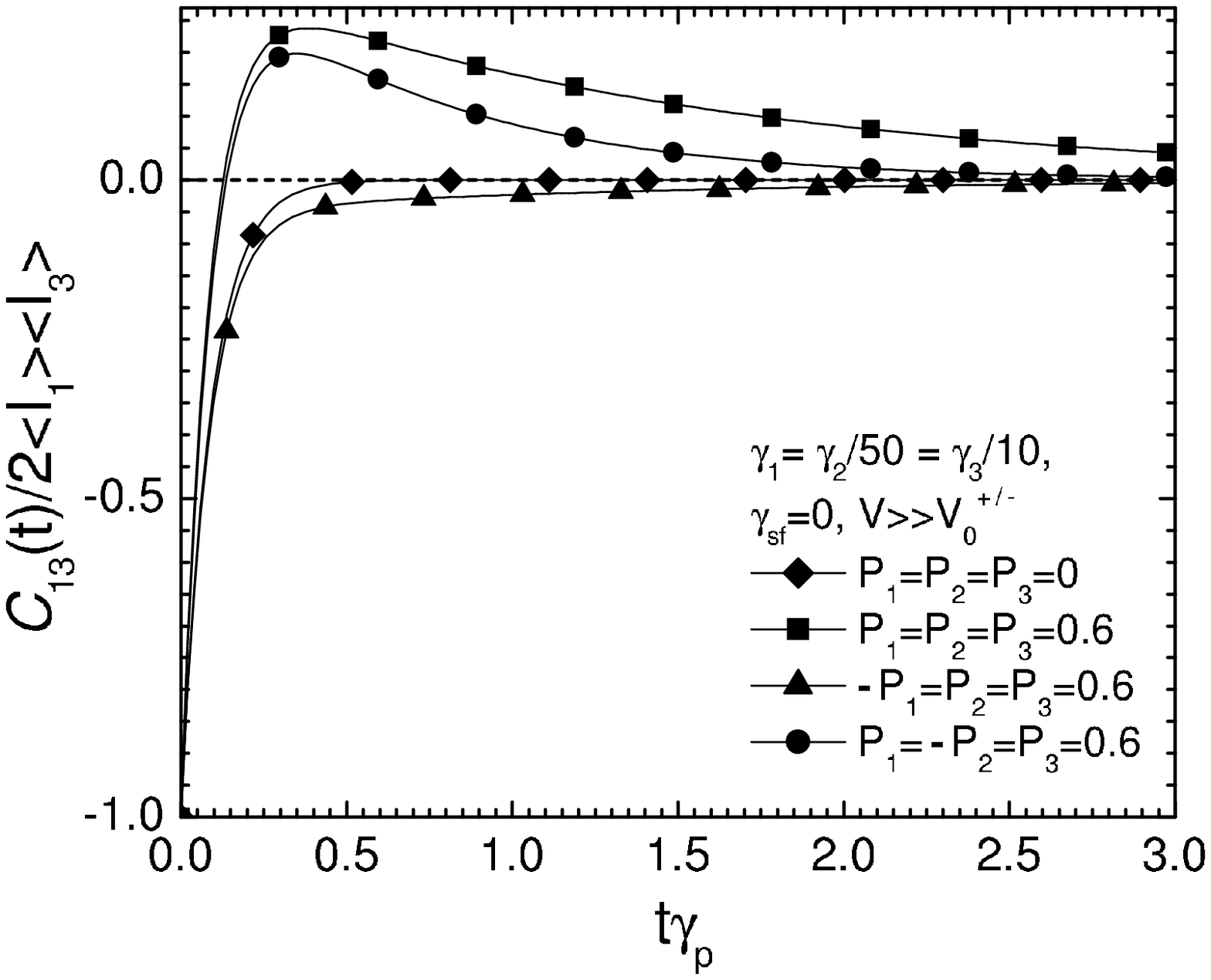}\caption{Time dependence of
$\mathcal{C}_{13}(t)$ in the limit $V\gg V_{0}^{sng(E_{0})}$, for the
different cases considered in Figs. \ref{Pict2} to \ref{Pict6}. Note that
$\mathcal{C}_{13}(t)$ is given in units of $-\mathcal{C}_{13}%
(t=0)=2\left\langle I_{1}\right\rangle \left\langle I_{3}\right\rangle $,
which depends on the polarization values.}%
\label{Pict8}%
\end{figure}

In this Section, we provide a physical explanation for the results of Sections
III.A and III.B, in the high-voltage limit $V\gg V_{0}^{sgn(E_{0})}$, where
the sign of $E_{0}$ does not matter. This analysis relies on the evaluation of
quantities defined in Section II.A and II.B (Table \ref{TableI}), on numerical
simulations of the temporal evolution of the spin $\sigma_{dot}$ of the dot
(Fig. \ref{Pict7}) and on plots of the correlation functions $\mathcal{C}%
_{13}(t)$ (Fig. \ref{Pict8}).

Let us first focus on the case $P_{1}=P_{2}=P_{3}=0.6$ (squares in Table
\ref{TableI}). For these values of lead polarization, up spins are in the
majority at the output. Thus, the dwell times of down spins on the dot is
longer than that of up spins ($t_{\downarrow}>t_{\uparrow}$ \ in Table
\ref{TableI}). However, one has $\bar{p}_{\downarrow}=\bar{p}_{\uparrow}$ thus
$\left\langle \sigma_{dot}\right\rangle =0$. This is because $t_{\downarrow
}>t_{\uparrow}$ is perfectly compensated by the fact that, due to $P_{2}>0$,
up electrons are in the majority in $I_{2}(t)$. Property $t_{\downarrow
}>t_{\uparrow}$ suggests that the up spins can flow through the dot only in
short time windows where the current transport is not blocked by a down spin.
This situation of \textquotedblleft dynamical spin-blockade\textquotedblright%
\ is responsible for a bunching of the tunneling events associated to the up
spins, as confirmed by the numerical simulations of $\sigma_{dot}(t)$ (Fig.
\ref{Pict7}). The average number of up spins grouped in a \textquotedblleft
bunch\textquotedblright\ corresponds to the quantity $n_{b}$ given in Table
\ref{TableI} (see \cite{nb}). On the one hand, the phenomenon of up-spins
bunching is very strong since, here, $n_{b}=4$. On the other hand, one can see
that the positive sign of $S_{13}(\omega=0)$ stems from the up-up correlations
(see $S_{1\uparrow,3\uparrow}(\omega=0)>0$ in Table \ref{TableI}). Therefore,
one question to answer is whether the positive sign of $S_{13}(\omega=0)$ can
be attributed to this bunching of up-spins tunneling events. For that purpose,
we have plotted the correlation function $\mathcal{C}_{13}(t)$ (Fig.
\ref{Pict8}) and compared it to the characteristic time scales of the
electronic transport. The correlation function $\mathcal{C}_{13}(t)$ is
negative for times shorter than (approximately) the average delay $t_{0}$
between the occupancy of the dot by two consecutive electrons. Then,
$\mathcal{C}_{13}(t)$ becomes positive and reaches a maximum at a time
comparable to the average delay $t_{0}+t_{\uparrow}$ between the passage of
two up spins on the dot. Eventually, $\mathcal{C}_{13}(t)$ is strongly
decreased at times of the order of the average duration $t_{b}$ of the
\textquotedblleft bunch\textquotedblright\ of spins. Hence, the
time-dependence of $\mathcal{C}_{13}(t)$ allows us to attribute the positive
value of $S_{13}(\omega=0)$ to the bunching of tunneling events caused by
dynamical spin-blockade. The same reasoning can be made to explain the
super-Poissonian value of $F_{2}$ (data not shown).

In the case $-P_{1}=P_{2}=P_{3}=0.6$ (triangles in Table \ref{TableI}), the
temporal evolution of $\sigma_{dot}$ (see Fig. \ref{Pict7}) is qualitatively
similar to that of the case $P_{1}=P_{2}=P_{3}=0.6$, thus up-up correlations
caused by dynamical spin-blockade lead again to a super-Poissonian $F_{2}$.
However, less up electrons flow through lead 1 than in the previous case
because the polarization $P_{1}$ has been reversed (see $I_{1\uparrow
}/I_{3\uparrow}$ in Table \ref{TableI}). Hence, the positive term
$S_{1\uparrow,3\uparrow}(\omega=0)$ is not large enough to lead to a positive
$S_{13}(\omega=0)$.

In the case $P_{1}=-P_{2}=P_{3}=0.6$ (circles in Table \ref{TableI}), there is
still dynamical spin-blockade, as shown by $t_{\downarrow}>t_{\uparrow}$ in
Table \ref{TableI}. This dynamical spin-blockade induces again a bunching of
the tunneling of up spins (see $S_{1\uparrow,3\uparrow}(\omega=0)>0$ in Table
\ref{TableI}). However, the up-up correlations are much weaker than in the
$P_{1}=P_{2}=P_{3}=0.6$ case due to the minority of up spins at the input.
Another positive contribution to the cross-correlations stems from the up-down
terms (see $S_{1\sigma,3-\sigma}(\omega=0)>0$ in Table \ref{TableI}). In fact,
since the average number $n_{b}=0.25$ of up spins passing consecutively
through the dot is very low \cite{nb}, we have $t_{b}<t_{\downarrow}$. Then,
each up spin is positively correlated to the first down spin preceding him
(see Fig. \ref{Pict8}). As a result, dynamical spin-blockade now produces a
bunching of tunneling events responsible for up-up \textit{and} up-down
correlations. The correlation function $\mathcal{C}_{13}(t)$ differs from that
of the case $P_{1}=P_{2}=P_{3}=0.6$ in the sense that it decreases more
quickly after its maximum, due to the smaller value of $t_{b}$. However,
contrarily to the case $P_{1}=P_{2}=P_{3}=0.6$, the decay time of
$\mathcal{C}_{13}(t)$ is much larger than $t_{b}$, due to fluctuations in the
number of spins per bunch with respect to $n_{b}=0.25$ (Fig. \ref{Pict7}).

In conclusion, we have seen that in all the cases treated here, the
super-Poissonian value of $F_{2}$ and the positive sign of $F_{13}$ can be
explained from the dynamical spin-blockade mechanism which induces a bunching
of the tunneling events.

\subsection{D. Effect of tunneling asymmetry}

\begin{figure}[h]
\includegraphics[width=1.\linewidth]{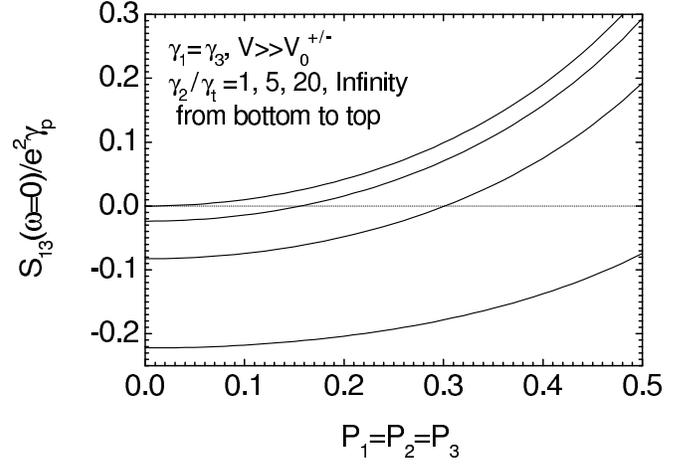}\caption{Influence of the
asymmetry between $\gamma_{2}$ and $\gamma_{t}$ on $S_{13}\left(
\omega=0\right)  $, for $V\gg V_{0}^{sgn(E_{0})}$, $P_{1}=P_{3}=P_{3}$,
$\gamma_{1}=\gamma_{3}$ and $\gamma_{sf}=0$. According to (\ref{Sexplicit}),
$S_{13}\left(  \omega=0\right)  $ is always positive for high enough values of
$P_{1}$. Large ratios $\gamma_{2}/\widetilde{\gamma}$ favor positive
cross-correlations by extending the positivity domain to lower polarization
values. In the limit $\gamma_{2}\gg\gamma_{t}$, for $P_{1}=P_{2}=P_{3}$, one
has $S_{13}\left(  \omega=0\right)  =4\gamma_{1}\gamma_{3}P_{1}^{2}%
e^{2}/\gamma_{t}\left(  1-P_{1}^{2}\right)  $ independent of $\gamma_{2}$.
These high-voltage results are independent of the values of $C_{i}$
considered.}%
\label{Pict9}%
\end{figure}

We now address the problem of how to choose parameters that favor the
observation of positive cross-correlations in the ferromagnetic case treated
here. First, from Section II.C, finite lead polarizations are necessary.
However, it is possible to get positive cross-correlations even if $P_{2}=0$,
provided the output of the device is sufficiently polarized. For instance, in
the high-voltage limit $V\gg V_{0}^{sgn(E_{0})},$ choosing $P_{1}=P_{3}$,
$P_{2}=0$ and $\gamma_{sf}=0$ leads to
\begin{equation}
S_{13}\left(  \omega=0\right)  =\frac{16e^{2}\gamma_{1}\gamma_{2}^{2}%
\gamma_{3}\left(  1-P_{1}^{2}\right)  [P_{1}^{2}(2\gamma_{2}+\gamma
_{t})-\gamma_{t}]}{\gamma_{t}[2\gamma_{2}+\left(  1-P_{1}^{2}\right)
\gamma_{t}]^{3}}\;\text{.}%
\end{equation}
In this limit, the current $\langle I_{2}\rangle$ \textit{is not spin
polarized}, i.e. $\left\langle I_{2,\uparrow}\right\rangle =\left\langle
I_{2,\downarrow}\right\rangle $, because up and down spins have the same
probability to enter the dot, regardless of what happens at the output. The
case where the three electrodes are polarized in the same direction leads to a
higher positive $S_{13}\left(  \omega=0\right)  $. Indeed, in the high-voltage
limit, choosing $P_{1}=P_{2}=P_{3}$ and $\gamma_{sf}=0$ leads to
\begin{equation}
S_{13}\left(  \omega=0\right)  =\frac{16e^{2}\gamma_{1}\gamma_{2}^{2}%
\gamma_{3}[P_{1}^{2}(2\gamma_{2}+\gamma_{t})-\gamma_{t}]}{\gamma_{t}%
(2\gamma_{2}+\gamma_{t})^{3}(1-P_{1}^{2})}\;\text{.} \label{Sexplicit}%
\end{equation}
The asymmetry between the tunneling rates $\gamma_{i}$ has a strong influence
on the cross-correlations. From (\ref{Sexplicit}), the case of symmetric
output junctions, i.e. $\gamma_{1}=\gamma_{3},$ is the most favorable
configuration for getting a large $S_{13}\left(  \omega=0\right)  >0$
\cite{choice}. In addition, choosing large values of $\gamma_{2}/\gamma_{t}$
decreases $\bar{p}_{0}$, which allows to extend the domain of positive
cross-correlations to smaller values of polarizations (Fig.~\ref{Pict9}). This
is important because ferromagnetic materials are usually not fully polarized
\cite{Soulen}.

\subsection{E. Finite frequency results}

\begin{figure}[h]
\includegraphics[width=1.\linewidth]{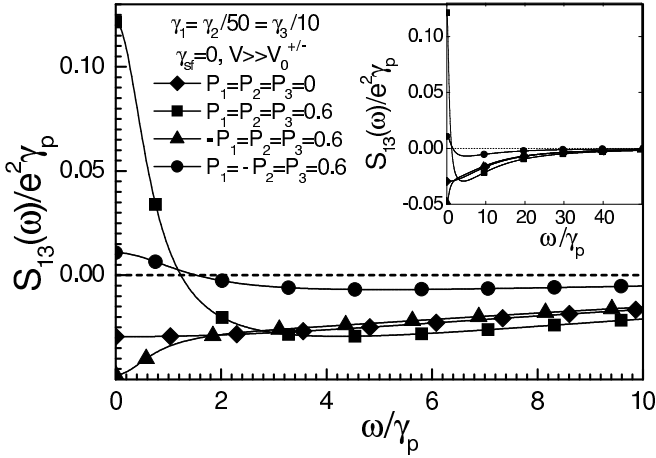}\caption{Frequency dependence
of $S_{13}\left(  \omega\right)  $ in the high-voltage limit $V\gg
V_{0}^{sgn(E_{0})}$, for the different cases considered in Figs. \ref{Pict2}
to \ref{Pict6}. The inset shows the same data for a larger frequency scale.}%
\label{Pict10}%
\end{figure}

Equation~(\ref{resultSij}) gives the frequency dependence of $S_{13}(\omega)$.
The spectrum $S_{13}(\omega)$ deviates from its zero frequency limit for
$\omega\gtrsim\left\vert \lambda_{+}\right\vert $. In the case $S_{13}%
(\omega=0)>0$, properties $\left\vert \lambda_{+}\right\vert <\left\vert
\lambda_{-}\right\vert $, $\mathcal{C}_{13}(t=0)<0$, $S_{13}^{+}>0$ and
$S_{13}^{-}<0$ (see Section II.C) imply that cross-correlations always turn to
negative when $\omega$ increases. Then, for frequencies larger than
$\left\vert \lambda_{-}\right\vert $, $S_{13}(\omega)$ tends to zero (see Fig.
\ref{Pict10}).

Eq. (\ref{Sijtot}) gives the expression of the total current
cross-correlations $S_{13}^{tot}(\omega)$ measured in practice, including the
contribution of screening currents. The spectrum $S_{13}^{tot}(\omega)$
differs from $S_{13}(\omega)$ only for $\omega\gtrsim\left\vert \lambda
_{+}\right\vert $. At large frequencies\ $\omega\gg\left\vert \lambda
_{-}\right\vert $, $S_{13}^{tot}(\omega)$ become a linear mixture of the
Schottky noises through the three junctions. If we furthermore assume $V\gg
V_{0}^{sgn(E_{0})}$, current conservation leads to%

\[
\frac{S_{13}^{tot}(\omega)}{2e}=\frac{I_{1}C_{3}}{C^{2}}\left(  C_{1}%
-C_{2}-C_{3}\right)  +\frac{I_{3}C_{1}}{C^{2}}\left(  C_{3}-C_{1}%
-C_{2}\right)  \text{ .}%
\]
This limit depends on the values of $C_{i}$ considered, in contrast to what
happens for $S_{ij}(\omega)$. It can be positive as well as negative depending
on the values of parameters.\ For $P_{1}=P_{2}=P_{3}=0.6$, $\gamma_{1}%
=\gamma_{2}/50=\gamma_{3}/10$, $C_{1}=C_{2}=C_{3}$ and $V\gg V_{0}%
^{sgn(E_{0})}$, one has a cross-over from positive to negative
cross-correlations as $\omega$ increases ($S_{13}^{tot}(\omega=0)/e^{2}%
\gamma_{p}\simeq+0.121$ and $S_{13}^{tot}(\omega\gg\left\vert \lambda
_{-}\right\vert )/e^{2}\gamma_{p}\simeq-0.222$). But the opposite situation is
also possible. For instance, with $P_{1}=P_{2}=P_{3}=0$, $\gamma_{1}%
=\gamma_{2}/50=\gamma_{3}/10$, $C_{1}=C_{2}=C_{3}/5$ and $V\gg V_{0}%
^{sgn(E_{0})}$, one has $S_{13}^{tot}(\omega=0)/e^{2}\gamma_{p}\simeq-0.030<0$
and $S_{13}^{tot}(\omega\gg\left\vert \lambda_{-}\right\vert )/e^{2}\gamma
_{p}\simeq+0.019>0$. For other positive cross-correlations due to screening
currents, see \cite{martin:00}. We recall that the results shown in this
Section are valid if the Markovian approximation holds, i.e. here $\hbar
\omega<\min\limits_{i}(\left\vert E_{0}-eV_{i}\right\vert )$ \cite{Engel}. The
results for the correlations of $I_{j}^{tot}(t)$ are furthermore valid only
for $\omega$ larger than the characteristic frequencies associated to the
charging of the capacitors (see Section II.E).

\subsection{F. Comments}

In spite of the large variety of proposals for getting positive
cross-correlations, this effect has not been observed experimentally yet. We
believe that the mechanism proposed in Section III can be implemented with
present techniques. For $\gamma_{1}=\gamma_{2}/10=\gamma_{3}$, the
polarizations $P_{1}=P_{2}=P_{3}=0.4$ typical for Co \cite{Soulen} lead to
positive cross-correlations of the order of $S_{13}/e^{2}\gamma_{tot}%
\simeq0.08$. With $\gamma_{p}\simeq5~$GHz, this corresponds to a current noise
level of $10^{-29}$A$^{2}$s. The maximum differential conductance of the
sample depends on temperature: $d\left\langle I_{2}\right\rangle /dV\sim
e^{2}\gamma_{p}\left(  C_{1}+C_{3}\right)  /5k_{B}TC$. Assuming that $T=20$ mK
and $C_{1}=C_{2}=C_{3}$, one obtains $\left(  d\left\langle I_{2}\right\rangle
/dV\right)  ^{-1}\sim h/e^{2}$. This leads to a voltage noise level measurable
with existing voltage noise-amplification techniques \cite{birk:95,Glattli}.

One difficulty of this experiment is connecting three leads to a very small
structure. We believe that a multiwall carbon nanotube (MWNT) contacted by
ferromagnetic leads could be an interesting candidate for implementing a
three-terminal device. The question of whether a MWNT splits into two quantum
dots when three contacts are evaporated on top it is still open. However,
given that the intrinsic level spacing of a MWNT connected to two leads seems
to be determined by its total length rather than by the separation between the
leads \cite{Buitelaar}, a three-terminal quantum dot structure seems feasible.
In addition, it has been demonstrated experimentally that contacting
ferromagnetic leads to a MWNT is possible \cite{TubesFerro}.

Interestingly, a different mechanism, proposed by Sauret and Feinberg, can
also lead to positive current cross-correlations in a quantum-dot circuit
\cite{sauret:03}. This work also considers current transport through one
single orbital of the dot. For certain bias voltages large enough to allow a
double occupation of this orbital, the Pauli principle induces positive
correlations between up and down spins. This so-called mechanism of
\textquotedblleft opposite-spin bunching\textquotedblright\ is antagonist to
our mechanism of dynamical spin-blockade which requires that the orbital can
be only singly occupied. However, with both mechanisms, positive
cross-correlations can be obtained only when the two spin channels do not
transport current independently, i.e. when charging effects are relevant
\cite{TwoIndep}. We point out that in the three-terminal geometry of Figure
(\ref{Pict1}), the opposite-spin bunching proposed by Sauret \textit{et al.}
allows to get positive output cross-correlations in spite of a sub-Poissonian
input Fano factor. This feature, added to our findings, shows that positive
output cross-correlations and a super-Poissonian input Fano factor can be
obtained \textit{separately} for a quantum dot connected to ferromagnetic
leads. Nevertheless, the opposite-spin bunching proposed by Sauret \textit{et
al.} can lead to positive cross-correlations between the total currents
through leads 1 and 3 only when the output leads are \textit{strongly}
polarized in opposite directions, in order to filter the weak up-down positive
cross-correlations induced by this effect. In practice, this is very difficult
to achieve with usual ferromagnetic materials \cite{Soulen}.

Note that the dynamical spin-blockade studied in this article is unrelated
with another mechanism called \textquotedblleft
spin-blockade\textquotedblright, observed in many semiconductor quantum dots
experiments (see \cite{Huttel} and references therein). This other
spin-blockade refers to the suppression of peaks expected in the I-V
characteristics of a quantum dot for independent single electron states, but
not observed due to quantum mechanical spin selection rules.

\section{IV. One-orbital quantum dot in a magnetic field, connected to three
paramagnetic leads}

In view of the experimental difficulties for connecting ferromagnetic leads to
semiconductor quantum dots \cite{Schmidt}, the question of whether it is
possible to obtain positive cross-correlations without using ferromagnetic
leads is of great interest. We thus consider in Sections IV.A and IV.B the
one-orbital case introduced in section II, with $P_{1}=P_{2}=P_{3}=0$ and
$B\neq0$.

At $B\neq0$, two resonances are expected a priori in the voltage
characteristics, for $V\simeq V_{\uparrow}^{sgn(E_{0})}$ and $V\simeq
V_{\downarrow}^{sgn(E_{0})}$. The limit $V\gg V_{\max}^{sgn(E_{0})}$ and
$\gamma_{sf}=0$ is the same as in the $B=0$ case because the tunneling rates
saturate at high voltages. In particular, from Eqs. (\ref{SFrelation}),
(\ref{curPara}) and (\ref{Sexplicit}), we have in this limit
\begin{align}
F_{2}  &  =\frac{4\gamma_{2}^{2}+\gamma_{t}^{2}}{\left(  \gamma_{t}%
+2\gamma_{2}\right)  ^{2}}\nonumber\\
F_{13}  &  =-\frac{4\gamma_{1}\gamma_{2}\gamma_{3}}{\gamma_{t}\left(
\gamma_{t}+2\gamma_{2}\right)  ^{2}}\text{ .} \label{INDEPcross}%
\end{align}
This means that here, a super-Poissonian $F_{2}$ and positive
cross-correlations can appear only at lower voltages, for which the cases
$E_{0}>0$ and $E_{0}<0$ differ significantly. Note that due to $P_{1}%
=P_{2}=P_{3}=0$, one obtains from Eqs. (\ref{MatrixM}) and (\ref{eqnoise3}):
\begin{equation}
S_{13}=\frac{\gamma_{1}\gamma_{3}}{\gamma_{t}}\mathcal{F}(\frac{\gamma_{2}%
}{\gamma_{t}},\frac{\gamma_{\uparrow\downarrow}}{\gamma_{t}},\frac
{\gamma_{\downarrow\uparrow}}{\gamma_{t}},\frac{E_{0}}{T},\frac{V}{T},\frac
{B}{T}) \label{rel}%
\end{equation}
According to (\ref{rel}), for a constant value of $\gamma_{t}$, $\gamma
_{1}=\gamma_{3}$ allows to maximize $\left\vert S_{13}\right\vert $.
Therefore, in this section, we will plot curves for $\gamma_{1}=\gamma_{3}$.
\begin{figure}[h]
\includegraphics[width=1.\linewidth]{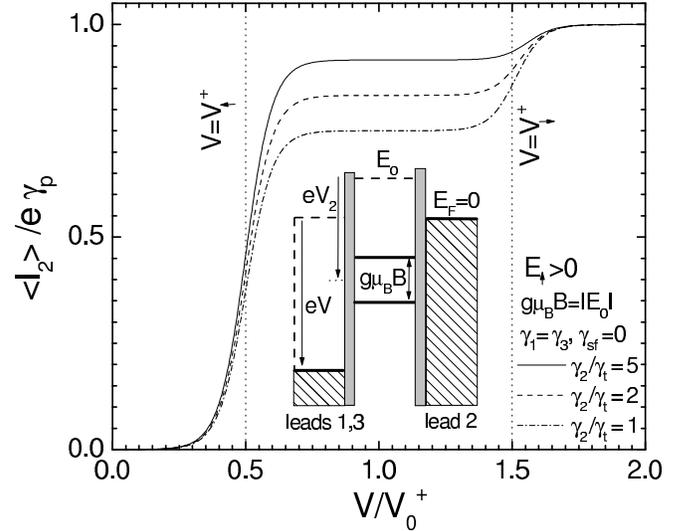}\caption{Average current
$\left\langle I_{2}\right\rangle $ as a function of the bias voltage $V$ for
$E_{0}>0$, $P_{1}=P_{2}=P_{3}=0$ , $C_{1}=C_{2}=C_{3}$, $\gamma_{1}=\gamma
_{3}$, $k_{B}T/\left\vert E_{0}\right\vert =0.05$, $g\mu_{B}B/\left\vert
E_{0}\right\vert =1$, and different values of $\gamma_{2}/\gamma_{t}$. These
curves display two steps, for $V\simeq V_{\uparrow}^{+}$ and $V\simeq
V_{\downarrow}^{+}$. The inset shows electrochemical potentials in the
circuit.}%
\label{Pict11}%
\end{figure}\begin{figure}[hh]
\includegraphics[width=1.\linewidth]{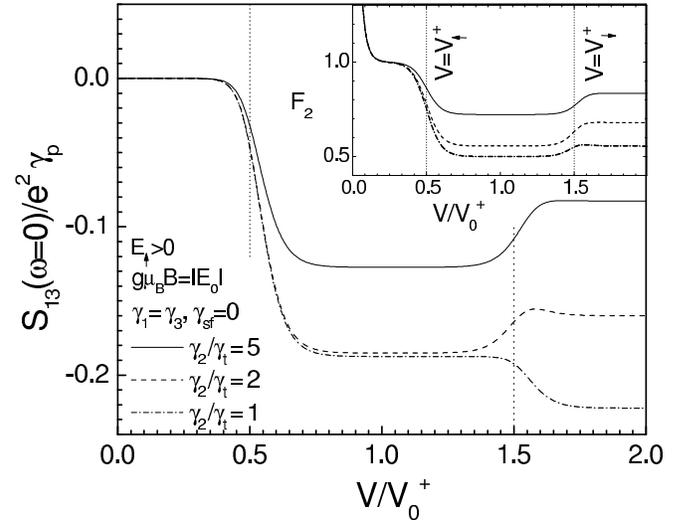}\caption{Zero-frequency
current cross-correlations $S_{13}(\omega=0)$ between leads $1$ and $3$ as a
function of the bias voltage $V$ for the same circuit parameters as in Fig.
\ref{Pict11}. Inset: Fano factor $F_{2}$. These curves display two steps, for
$V\simeq V_{\uparrow}^{+}$ and $V\simeq V_{\downarrow}^{+}$. Above the thermal
peak, for $\gamma_{sf}=0$, one has $F_{2}\leq1$ and $S_{13}(\omega=0)\leq0$
for any values of the parameters.}%
\label{Pict12}%
\end{figure}

\subsection{A. Zero-frequency results for $E_{0}>0$}

We first briefly comment the case in which the two Zeeman sublevels are above
the Fermi energy at equilibrium (\textit{i.~e}. $E_{\uparrow(\downarrow)}>0$).
The current and noise voltage-characteristics obtained in this situation were
already discussed in Ref.~\cite{thielmann:03} for the two-terminal case. Like
in III.A, for $V<V_{\uparrow}^{+}$, $\left\langle I_{2}\right\rangle $ and
$S_{13}(\omega=0)$ are exponentially small and $F_{2}$ is Poissonian with a
thermal peak at $V\rightarrow0$, followed by a unitary plateau (Figs.
\ref{Pict11} and \ref{Pict12}). Then, the curves $\left\langle I_{2}%
\right\rangle ,$ $F_{2}$ and $S_{13}(\omega=0)$ show two steps corresponding
to $V\simeq V_{\uparrow}^{+}$ and then $V\simeq V_{\downarrow}^{+}$. We have
verified analytically that, above the thermal peak, for $\gamma_{sf}=0$, one
has $F_{2}\leq1$ and $F_{13}\leq0$ for any values of the parameters. For
$V<V_{\downarrow}^{+}$, the current $\langle I_{2}\rangle$ is spin polarized,
an effect which allows to do spin filtering with a nearly 100\% efficiency
\cite{Recher,hanson:03a}.

\subsection{B. Zero-frequency results for $E_{0}<0$}

\begin{figure}[h]
\includegraphics[width=1.\linewidth]{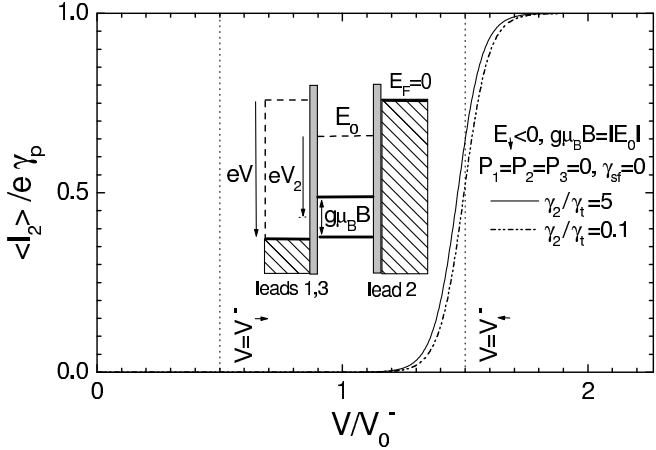}\caption{Current-voltage
characteristic of the circuit of Fig. \ref{Pict1} for $E_{0}<0$, $P_{1}%
=P_{2}=P_{3}=0$ , $C_{1}=C_{2}=C_{3}$, $\gamma_{1}=\gamma_{3}$, $k_{B}%
T/\left\vert E_{0}\right\vert =0.05$, $g\mu_{B}B/\left\vert E_{0}\right\vert
=1$, and different values of the asymmetry $\gamma_{2}/\gamma_{t}$ between the
input and the output. These curves display only one step, for $V\simeq
V_{\uparrow}^{-} $. The inset shows the electrochemical potentials in the
circuit.}%
\label{Pict13}%
\end{figure}\begin{figure}[hh]
\includegraphics[width=1.\linewidth]{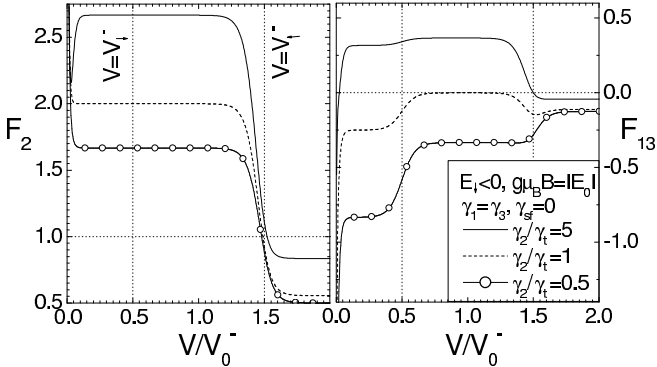}\caption{Input Fano factor
$F_{2}$ (left panel) and output cross-Fano factor $F_{13}$ (right panel) as a
function of the bias voltage $V$, for the same circuit parameters as in Fig.
\ref{Pict13}. In all curves $\gamma_{sf}=0.$ $F_{2}$ can be super-Poissonian
and $F_{13}$ positive for certain values of $\gamma_{2}/\gamma_{t}$ (see
text). The Fano factor $F_{2}$ shows only one step for $V\simeq V_{\uparrow
}^{-}$ whereas $F_{13}$ shows two steps, for $V\simeq V_{\downarrow}^{-}$ and
$V\simeq V_{\uparrow}^{-}$.}%
\label{Pict14}%
\end{figure}

Below, we focus on the case in which the two Zeeman sublevels are below the
Fermi energy at equilibrium (i.e. $E_{\uparrow(\downarrow)}<0$). To our
knowledge, the current noise in this configuration has never been studied
before, even for a two-terminal device. We will first study analytically what
happens above the thermal peak, i.e. $eV\gg k_{B}T$. In this limit, one can
write the tunneling rates as $\Gamma_{2\sigma}^{+}=\gamma_{2}$, $\Gamma
_{2\sigma}^{-}=0$, $\Gamma_{1(3)\uparrow}^{-}=x\gamma_{1(3)}$, $\Gamma
_{1(3)\uparrow}^{+}=(1-x)\gamma_{1(3)}$, $\Gamma_{1(3)\downarrow}^{-}%
=y\gamma_{1(3)}$, and $\Gamma_{1(3)\downarrow}^{+}=(1-y)\gamma_{1(3)}$, where
$x=1/(1+\exp[-(E_{\uparrow}-eV_{1})/k_{B}T])$ and $y=1/(1+\exp[-(E_{\downarrow
}-eV_{1})/k_{B}T])$. The hypothesis $B>0$ implies that $x<y$. First, for
$V<V_{\downarrow}^{-},$ we have $x\rightarrow0$ and $y\rightarrow0$. \ Then,
the parameters $x$ and $y$ go from 0 to 1 while the voltage increases. For
$V=V_{\downarrow}^{-}$ i.e. $y=1/2$, the upper Zeeman sublevel is at resonance
with the Fermi level of the output leads 1 and 3. Then, for $V=V_{\uparrow
}^{-}$ i.e. $x=1/2$, the lower Zeeman sublevel is at resonance with the
outputs, as represented by the level diagram in Fig.~\ref{Pict13}.

The assumptions made on the rates lead to
\begin{equation}
F_{13}=\frac{\gamma_{1}\gamma_{3}}{\gamma_{2}\gamma_{t}^{2}}\left[  \gamma
_{2}\left(  F_{2}-1\right)  +\gamma_{t}(x+y-2)\right]  \text{ .}
\label{F13para}%
\end{equation}
In Section II, we have shown that relation (\ref{SFrelation}) between $F_{2}$
and $F_{13}$ is\ always valid at high voltages (i.e $x,y\sim1$ here) for the
single-orbital problem with $P_{1}=P_{2}=P_{3}$. But this demonstration does
not take into account the symmetries that the problem takes for certain
particular cases. Here, from (\ref{F13para}), $P_{1}=P_{2}=P_{3}=0$ implies
that property (\ref{SFrelation}) is also valid at any $V$ above the thermal
peak when $\gamma_{2}\gg\gamma_{t}$.

The inequality $t_{\downarrow}=1/y\gamma_{t}\neq t_{\uparrow}=1/x\gamma_{t}$
for $x\neq y$ suggests the possibility of obtaining again dynamical
spin-blockade. To study the situation accurately, we will consider the
simplified situation $k_{B}T\ll g\mu_{B}B$, i.e. the up-spins channel starts
to conduct for voltages such that down spin can flow only from the right to
the left. This means that for the first voltage transition $V\simeq
V_{\downarrow}^{-}$, we have $x\ll1$ and it is enough to consider low order
developments of $\langle I_{2}\rangle$, $F_{2}$, and $F_{13}$ with respect to
$x$:
\begin{equation}
\langle I_{2}\rangle=\frac{2e\gamma_{2}\gamma_{t}x}{\tilde{\gamma}+\gamma_{2}%
}\,+o(x)^{2}\text{ ,} \label{I2dev}%
\end{equation}%
\begin{equation}
F_{2}=\frac{\gamma_{t}+3\gamma_{2}}{\gamma_{t}+\gamma_{2}}+o(x)\text{ }
\label{F2dev}%
\end{equation}
and
\begin{equation}
F_{13}=\frac{\gamma_{1}\gamma_{3}\left[  (2\gamma_{2}^{2}-\gamma_{t}\left(
\gamma_{t}+\gamma_{2}\right)  \left(  2-y\right)  \right]  }{\left(
\gamma_{t}+\gamma_{2}\right)  \gamma_{2}\gamma_{t}^{2}}+o(x)\text{ }
\label{S13dev}%
\end{equation}
for $\gamma_{sf}=0$. Transport through the upper level is energetically
allowed for $y>1/2$. However, since we have assumed $x\ll1$, from Eq.
(\ref{I2dev}), $\langle I_{2}\rangle$ remains very small throughout the
$V\simeq V_{\downarrow}^{-}$\ transition: the dot is blocked by up spins, thus
down spins cannot cross the dot. Even if the current is very low, this leads
to dynamical spin-blockade and thus to a super-Poissonian $F_{2}$, except in
the limit $\gamma_{t}\gg\gamma_{2}$ [see Eq. (\ref{F2dev})]. Accordingly,
$F_{13}$ can be positive for certain tunneling rate asymmetries [Eq.
(\ref{S13dev})]. The factor $F_{13}$ shows a step around $V\simeq
V_{\downarrow}^{-}$, due to the $y$ dependence in Eq. (\ref{S13dev}), whereas
$F_{2}$ is constant throughout the $V\simeq V_{\downarrow}^{-}$ transition.
This implies a redistribution of the zero-frequency noise between
$S_{11}(\omega=0)$, $S_{33}(\omega=0)$ and $S_{13}(\omega=0)$ when the
threshold $V=V_{\downarrow}^{-}$ is crossed [see (\ref{sum2})]. The absence of
step for $F_{2}$ can be attributed to the unidirectionality of tunneling
through junction 2. Indeed, $x\rightarrow0$ means that $F_{2}$ depends only on
$p_{0}$ and $G_{0,\uparrow(\downarrow)}$ [see (\ref{curpart}) and
(\ref{eqnoise3})]. Now, for $V\simeq V_{\downarrow}^{-},$ the contribution of
these terms is independent of $y$ (and thus on $V$) at first order in $x$,
because $\bar{p}_{0}$ and $G_{0,\uparrow(\downarrow)}$ are already forced to
very low values due to the $x\rightarrow0$ hypothesis. On the contrary,
$F_{13}$ also depends on $\bar{p}_{\uparrow,\downarrow}$ and $G_{\sigma,0}$
with $\sigma\in\left\{  \uparrow,\downarrow,0\right\}  $. For $x\rightarrow0$,
these last terms depend strongly on $y$.\ \begin{figure}[h]
\includegraphics[width=1.\linewidth]{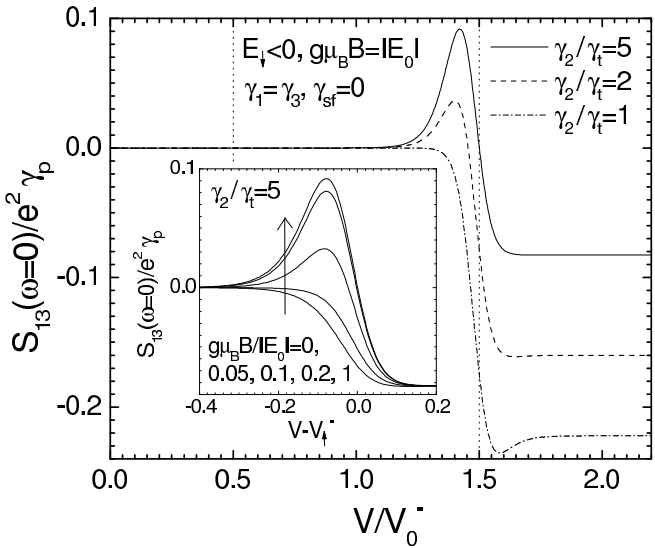}\caption{Zero-frequency
current cross-correlations $S_{13}(\omega=0)$ between leads $1$ and $3$ as a
function of the bias voltage $V$, for the same circuit parameters as in Fig.
\ref{Pict13} and different values of junction asymmetry. The inset shows the
effect of a magnetic field $B$ for $\gamma_{2}/\gamma_{t}=5$ and $\gamma
_{sf}=0$. }%
\label{Pict15}%
\end{figure}\begin{figure}[hh]
\includegraphics[width=1.\linewidth]{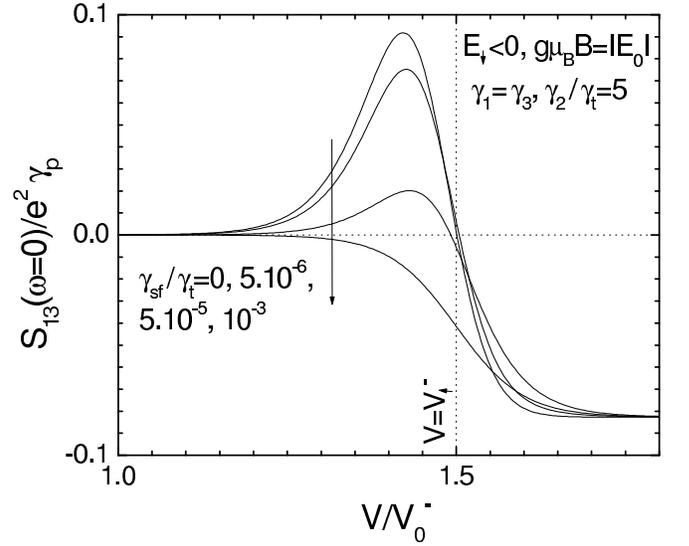}\caption{Effect of spin
flip-scattering on the current cross-correlations between leads $1$ and $3$
for the same circuit parameters as in Fig. \ref{Pict15} and $\gamma_{2}%
/\gamma_{t}=5$.}%
\label{Pict15bis}%
\end{figure}

For $k_{B}T\ll g\mu_{B}B$, the second possible voltage transition $V\simeq
V_{\uparrow}^{-}$ can be described by taking the limit $y=1$ where%

\begin{equation}
\langle I_{2}\rangle=\frac{2ex\gamma_{2}\gamma_{t}}{\gamma_{t}+\gamma
_{2}(1+x)}\text{ ,} \label{I2para}%
\end{equation}%
\begin{equation}
F_{2}=1+\frac{2\gamma_{2}\left(  \gamma_{t}(1-3x)+(1-x)^{2}\gamma_{2}\right)
}{\left(  \gamma_{t}+\gamma_{2}(1+x)\right)  ^{2}} \label{F2para}%
\end{equation}
and%

\begin{multline}
F_{13}=\frac{\gamma_{1}\gamma_{3}}{\gamma_{t}^{2}}\frac{1}{\gamma_{2}%
(\gamma_{t}+(1+x)\gamma_{2})^{2}}[\label{eq:C2para}\\
2(1-x)^{2}\gamma_{2}^{3}+(1-7x+x^{2}+x^{3})\gamma_{2}^{2}\gamma_{t}\\
-2(1-x^{2})\gamma_{2}\gamma_{t}^{2}-(1-x)\gamma_{t}^{3}]\text{ }%
\end{multline}
for $\gamma_{sf}=0$. Around $V\simeq V_{\uparrow}^{-}$, the blockade of the
dot by up spins is partially lifted and transport through both levels is
allowed. The average input current $\langle I_{2}\rangle$ thus increases with
voltage (i.e. with $x$) [see (\ref{I2para})]. On the opposite of what happens
in IV.A, the average current $\langle I_{2}\rangle$ is \textit{not
spin-polarized} because up and down spin have the same probability to enter
the dot. The factors $F_{2}$ and $F_{13}$ both show a step through the
$V\simeq V_{\uparrow}^{-}$ transition (as indicated by their $x$-dependency)
and tend at high voltages to the usual sub-Poissonian values.

We now turn to the discussion of the general results displayed in
Figs.~(\ref{Pict13}-\ref{Pict15bis}), obtained from an exact treatment of the
full Master equation. Fig.~\ref{Pict13} shows the full voltage dependence of
$\langle I_{2}\rangle$. As expected from (\ref{I2dev}) and (\ref{I2para}),
this current shows a single step at $V\simeq V_{\uparrow}^{-}$, an effect
observed experimentally \cite{ralph,cobden,hanson:03}. The width on which
$\langle I_{2}\rangle$ varies is of the order of $\Delta V\sim10k_{B}TC/eC_{2}
$, whereas the position of the step varies only slightly with the asymmetry of
the junctions (the maximal variation is about $\Delta V^{\prime}\sim
0.7k_{B}TC/eC_{2}$).

The left panel of Fig. \ref{Pict14} shows the voltage dependence of $F_{2}$.
The divergence $2k_{B}T/eV$ of $F_{2}$ at zero voltage is again a result of
the dominating thermal noise in the limit $k_{B}T>eV$. As expected from
(\ref{F2dev}) and (\ref{F2para}), $F_{2}$ shows one single step at $V\simeq
V_{\uparrow}^{-}$, which is super-Poissonian except for $\gamma_{t}\gg
\gamma_{2}$. The right panel of Fig.~\ref{Pict14} shows the voltage dependence
of $F_{13}$. As expected, $F_{13}$ shows two steps at $V\simeq V_{\downarrow
}^{-}$ and then $V\simeq V_{\uparrow}^{-}$. The first plateau is positive for
$\gamma_{2}>\gamma_{t}\left(  1+\sqrt{5}\right)  /2$ and the second for
$\gamma_{2}>\gamma_{t}$, as can be seen from (\ref{S13dev}). The high-voltage
plateau is negative as usual. The case $\gamma_{2}/\gamma_{t}=1/2$ and
$V_{\downarrow}^{-}<V<V_{\uparrow}^{-}$ is one more illustration that it is
possible to have $F_{2}$ super-Poissonian and $F_{13}<0$.

It is also interesting to look at $S_{13}(\omega=0)$ which is the signal
measured in practice (Fig. \ref{Pict15}). Like $\left\langle I_{2}%
\right\rangle $, the cross-correlations $S_{13}(\omega=0)$ are exponentially
small for $V\simeq V_{\downarrow}^{-}$, thus the first voltage step of
$F_{13}$ is not visible on the scale of Fig. \ref{Pict15}. Cross-correlations
have a significant variation around $V\simeq V_{\uparrow}^{-}$, with a
positive peak for $\gamma_{2}>\gamma_{t}$. The maximum positive $S_{13}%
(\omega=0)$ obtained at this peak is of the same order as the maximum
$S_{13}(\omega=0)$ predicted in the ferromagnetic case for comparable junction
asymmetries (see Section III.F). Note that the height of the positive peak is
independent of temperature as long as (\ref{eq:parameters}) is fulfilled,
whereas its width, which is approximately $\Delta V$, depends on temperature.

Since the positive cross-correlations found in this work are due to dynamical
spin-blockade, we expect a strong dependence on the magnetic field. The inset
of Fig.~\ref{Pict15} shows the voltage dependence of $S_{13}(\omega=0)$ around
the step $V_{\uparrow}^{-}$, for a fixed temperature, a tunneling asymmetry
$\gamma_{2}/\gamma_{t}=5,$ and various magnetic fields. The amplitude of the
positive peak first increases with $B$ and then saturates once the Zeeman
splitting of the levels is much larger than the thermal smearing of the
resonances (i.e. $g\mu_{B}B\geq8k_{B}T$). The peak then simply shifts to
larger voltages while $B$ increases. Fig.~\ref{Pict15bis} shows the effect of
spin-flip scattering on the cross-correlations. Spin flips modify the positive
peak of $S_{13}(\omega=0)$ when $\Gamma_{\uparrow\downarrow}=\gamma_{sf}%
\exp(g\mu_{B}B/2k_{B}T)\sim\gamma_{i}$, see Eq.~(\ref{MatrixM}). It is thus
preferable to use a $B$ not larger than $8k_{B}T$ when spin flip scattering is
critical. As expected, a strong spin-flip scattering suppresses all
spin-effects and turns the positive cross-correlations to negative.

\subsection{C. Comments}

There is a strong qualitative difference between the ferromagnetic case of
Section III and the $B\neq0$ case of Section IV: in Section IV we have
obtained positive cross-correlations in the form of a peak around a resonance
voltage whereas in Section III, positive cross-correlations reach their
maximum above the resonance voltage.

In practice, we can imagine to tune the bias voltage $V$ such that different
orbital levels will transport current successively while the gate voltage of
the dot is swept, leading to an effective $E_{0}$ oscillating between positive
and negative values. In this situation, the results of Sections IV.A and V.B
indicate the possibility of having the sign of $S_{13}(\omega=0)$ which
oscillates with the gate voltage. \begin{figure}[h]
\includegraphics[width=1.\linewidth]{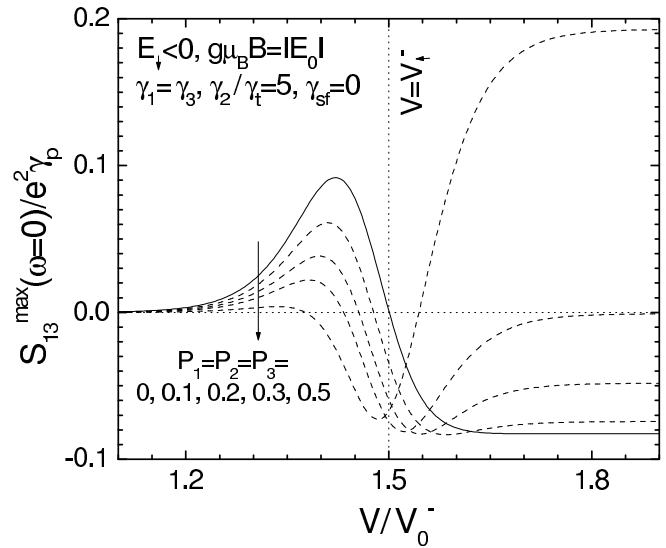}\caption{Zero-frequency
current cross-correlations between leads $1$ and $3$ as a function of the bias
voltage for the same circuit parameters as in Fig. \ref{Pict15}, $\gamma
_{2}/\gamma_{t}=5$ and $\gamma_{sf}=0$. The full line corresponds to the case
$P_{1}=P_{2}=P_{3}=0$ shown in Fig. \ref{Pict15} and the dashed lines to
finite values of $P_{1}=P_{2}=P_{3}$.}%
\label{Pict16}%
\end{figure}

MWNTs could be possible candidates for observing this effect. However, lateral
semiconductor quantum dots seem even more attractive. The fabrication
technology of lateral semiconductor quantum dots allows to engineer more than
two leads just by adjusting a lithography pattern (see for instance
\cite{Leturcq}). Another advantage of these structures is that the asymmetry
of the tunnel junctions, which is very critical for getting dynamical
spin-blockade, can be controlled just by changing the voltage of the gates
delimiting them. In addition, it has been shown that the spin-flip rate can be
very low in semiconductor quantum dots \cite{Khaestkii,hanson:03}. However,
implementing the model of Section IV requires that the leads can be considered
as unpolarized, which is not obvious in these systems if the magnetic field is
not applied locally to the dot but to the whole circuit. In certain cases, the
magnetic field can induce a significant spin polarization at the edges of the
two-dimensional electron gas, leading to different net tunneling rates
$\gamma_{j,\uparrow}$ and $\gamma_{j,\downarrow}$ for up and down spins
\cite{hanson:03a,Ciorga,Rogge}. In an extremely simplified approach, we have
taken this effect into account with finite polarizations $P_{1}=P_{2}=P_{3}$
with the same sign as $B$ (see Fig. \ref{Pict16}). The positive peak of
$S_{13}(\omega=0)$ is suppressed while $P_{1}$ increases because the tunneling
rates of spins which blocked the dot for $P_{1}=P_{2}=P_{3}=0$ increase.
However, this positive peak is replaced by a high-voltage positive limit
simply identical to that of Section III for the corresponding polarizations.
Note that for semiconductor quantum dots in the few electron regime, the time
evolution of $\left\vert \sigma_{dot}\right\vert $ can be measured by coupling
the dot to a single electron transistor or a quantum point contact
\cite{LU,Field,Hanson03c,ETHqpc}. In the high-voltage limit where current
transport is unidirectional, studying the statistics of $\left\vert
\sigma_{dot}(t)\right\vert $ would give a direct access to $S_{22}(\omega)$
for currents too low to be measured with standard techniques.

\section{V. Two-orbital spin-degenerate quantum dot circuit}

\subsection{A. Mapping onto the one-orbital non spin-degenerate case}

\begin{figure}[h]
\includegraphics[width=0.4\linewidth]{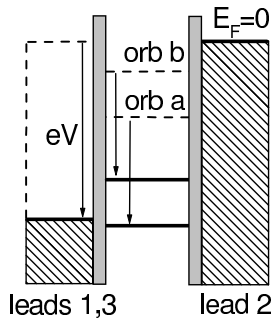}\caption{Electrochemical
potentials for a quantum dot connected to three paramagnetic leads and subject
to no magnetic field, with two different orbitals levels $a$ and $b$
accessible for current transport.}%
\label{Pict17}%
\end{figure}

We now consider the quantum dot circuit of Fig. \ref{Pict1} with $V>0$,
connected to paramagnetic leads ($P_{1}=P_{2}=P_{3}=0$) and with no magnetic
field ($B=0$). We assume that two different orbitals levels $a$ and $b$ of the
dot are accessible for current transport, but we still consider that the dot
cannot be doubly occupied. We define $\gamma_{j,orb}$ as the net tunneling
rate between lead $j$ and the orbital $orb\in\left\{  a,b\right\}  $. This
problem is spin-degenerate and can thus be treated without the spin degree of
freedom, which is replaced by the orbital degree of freedom. The rate for an
electron to tunnel between lead $j$ and the orbital level $orb$ in direction
$\epsilon$ is $\Gamma_{j,orb}^{\epsilon}=\gamma_{j,orb}/(1+\exp[\epsilon
(E_{orb}-eV_{j})/k_{B}T])$, where $E_{orb}$ is the intrinsic energy of the
orbital level $orb$. This problem can be treated in the sequential tunneling
limit with a Master equation analog to (\ref{MasterEquation}). There is in
fact a direct mapping between this problem and that described in Section II.
We will assume that $E_{a}<E_{b}$, so that the orbitals $a$ and $b$ will play
the roles of the Zeeman sublevels $\uparrow$ and $\downarrow$ of Section II,
where $B>0$. One has to replace the parameters of the previous problem by
\begin{gather}
E_{\uparrow(\downarrow)}\longrightarrow E_{a(b)}\text{ ,}\nonumber\\
\gamma_{j}\longrightarrow\frac{\gamma_{j,a}+\gamma_{j,b}}{2}\text{
,}\label{mapping}\\
P_{j}\longrightarrow\widetilde{P}_{j}=\frac{\gamma_{j,a}-\gamma_{j,b}}%
{\gamma_{j},_{a}+\gamma_{j,b}}\text{ ,}\nonumber\\
\gamma_{\uparrow\downarrow(\downarrow\uparrow)}\longrightarrow\gamma
_{ab(ba)}\text{ .}\nonumber
\end{gather}

This mapping shows that one can obtain positive cross-correlations in this
two-orbital system. It provides the evidence that interactions can lead to
zero-frequency positive cross-correlations in a normal quantum dot circuit
even \textit{without lifting spin-degeneracy. }Note that in practice,
$\gamma_{j,a}=\gamma_{j,b}$ is not obvious because of the different spatial
extensions of the orbitals (see for instance \cite{hanson:03a,Leturcq}). This
problem is thus equivalent to a one-orbital problem with $B\neq0$ \textit{and}
with finite effective polarizations $\widetilde{P}_{1},\widetilde{P}%
_{2},\widetilde{P}_{3}$ which can be close to $\pm1$. Positive
cross-correlations can be expected either at the resonance associated to level
$b$ (for $E_{b}<0$) or in the plateau following this resonance, depending on
the parameters (see for instance Fig. \ref{Pict16}).

\subsection{B. Comments}

In the one-orbital ferromagnetic case, we have shown that the simple relation
(\ref{SFrelation}) between $F_{2}$ and $F_{13}$ is valid in the high-voltage
limit only when $P_{1}=P_{3}$. Therefore, according to the mapping indicated
in Section V.A., in the two-orbital case, relation (\ref{SFrelation}) is valid
in the high-voltage limit only if $\widetilde{P}_{1}=\widetilde{P}_{3}$ i.e.
$\gamma_{1,a}/\gamma_{3,a}=\gamma_{1,b}/\gamma_{3,b}$. Hence, the range of
validity of property (\ref{SFrelation}) found in Section II.D for the
one-orbital system cannot be generalized to the two-orbital case.

In the spin-degenerate case treated here, positive cross-correlations stem
from\ the partial blockade of an electronic channel by another one, thus we
suggest to call this effect: \textit{dynamical channel-blockade. }This effect
should be observable in semiconductor quantum dots. The advantage of taking
$B=0$ is that the problem of spurious lead polarization evoked in Section IV
is suppressed. When $eV\gg\left\vert E_{b}+E_{C}\right\vert $ and $\gamma
_{sf}=0$, the two channels conduct current independently, thus dynamical
channel-blockade is suppressed and the positive cross-correlations disappear
(see (\ref{mapping})+\cite{TwoIndep}). When $\Delta E\ll k_{B}T$,
cross-correlations are always negative in a spin-degenerate three-terminal
quantum dot placed in the sequential tunneling limit \cite{bagrets:02}.
Therefore, the hypothesis $\Delta E\gg k_{B}T$ is also necessary to obtain
positive cross-correlations in this device. In fact, when $\Delta E\ll k_{B}T
$, the electron leaving the dot at a given time is not necessarily the one
which entered the dot just before, in spite of $eV\ll E_{C}$: channel effects
are suppressed.

Note that a super-Poissonian Fano factor can also be obtained in a
spin-degenerate circuit based on two bi-terminal quantum dots (or localized
impurity states) placed in parallel and coupled electrostatically to each
other \cite{Safonov,kiesslich,Nauen}. If one of the dots is charged, the other
cannot transport current because of the Coulomb repulsion. The dot which
changes its occupancy with a slower rate modulates the current through the
other one, which leads to a dynamical channel-blockade analogous to what we
found. The possibility to get positive cross-correlations in these systems was
not investigated, but Section V of the present article suggests it.

\section{VI. Conclusion}

We have considered noise in a three-terminal quantum dot operated as a beam
splitter. In this system, a super-Poissonian input Fano factor is not
equivalent to zero-frequency positive output cross-correlations. We have
studied three different ways to get these two effects, due to the mechanism of
dynamical channel-blockade.\ The first two strategies consist in involving
only one orbital of the dot in the electronic transport and lifting
spin-degeneracy, either by using ferromagnetic leads or by applying a magnetic
field to the dot. We have furthermore shown that lifting spin degeneracy is
not necessary anymore when two orbitals of the dot are involved in the current
transport. These results show that\textit{\ }one can get zero-frequency
positive cross-correlations due to interactions inside a beam splitter
circuit, even if this is a spin-degenerate normal fermionic circuit with a
perfect voltage-bias.

We thank H.~A. Engel, K.~Ensslin, M.~Governale, H.~Grabert, R.~Hanson,
T.~Kontos, R.~Leturcq, B.~Reulet, I.~Safi, P.~Samuelsson and B.~Trauzettel for
interesting discussions. We are particularly indebted to M.~B\"{u}ttiker for
raising the question which led us to consider the case treated in Section V.
This work was financially supported by the RTN Spintronics, by the Swiss NSF
and the NCCR Nanoscience.


\begin{thebibliography}{99}
\bibitem {nb}Note that $n_{b}=I_{2\uparrow}/I_{2\downarrow}$ includes bunches
with no up spins, i.e. two down spins passing consecutively. Thus, $n_{b}$ can
be smaller than $1$.
\bibitem {choice}Note that despite this, we have chosen to show in Sections
II.A and II.B curves for the case $\gamma_{3}/\gamma_{1}=10$ because it allows
to get, just by reversing the polarization of one output lead, both the cases
where $S_{13}\left(  \omega=0\right)  $ is positive or negative while $F_{2}$
is super-Poissonian.
\bibitem {TwoIndep}In the one-orbital case, positive cross-correlations can be
obtained only if the tunneling rates for one spin direction depend on the
occupation of the other spin level. When charging effects are not relevant,
i.e. $V\gg\left\vert E_{\uparrow}+E_{C}\right\vert C/\left(  C_{1}%
+C_{3}\right)  $ for $E_{0}>0$ or $V\gg\left\vert E_{\downarrow}%
+E_{C}\right\vert C/C_{2}$ for $E_{0}<0$, the tunneling rates become
independent of the occupation of the dot. We have verified analytically
(calculation not shown here) that in these conditions, for $\gamma_{sf}=0$,
$S_{13}\left(  \omega=0\right)  $ is always negative for any polarization of
the leads 1, 2, 3 and any spin-flip rate.
\end{thebibliography}
\end{document}